\documentclass[aps,prb,reprint,superscriptaddress]{revtex4-1}
\usepackage{amsfonts}
\usepackage{amssymb}
\usepackage{amsmath}
\usepackage{graphicx}
\usepackage{epsfig}
\usepackage{array}

\begin{document}

\title{Edge state in AB-stacked bilayer graphene and its correspondence with
SSH ladder}
\author{Tixuan Tan}
\email{ttx2000@hku.hk}
\author{Ci Li}
\email{oldsmith@hku.hk}
\author{Wang Yao}
\affiliation{Department of Physics, The University of Hong Kong, Hong Kong,
China}
\affiliation{HKU-UCAS Joint Institute of Theoretical and
Computational Physics at Hong Kong, China}
\begin{abstract}
We study edge states in AB-stacked bilayer graphene (BLG) ribbon where the
Chern number of the corresponding two-dimensional (2D) bulk Hamiltonian is
zero. The existence and topological features of edge states when two layers
ended with the same or different edge terminations (zigzag, bearded,
armchair) are discussed. The edge states (non-dispersive bands near the
Fermi level) are states localized at the edge of graphene nanoribbon that
only exists in certain range of momentum $k_y$. Their existence near the
Fermi level are protected by the chiral symmetry with topology well
described by coupled Su-Schrieffer-Heeger (SSH) chains model, i.e., SSH
ladder, based on the bulk-edge correspondence of one-dimensional (1D)
systems. These zero-energy edge states can exist in the whole $k_y$ region
when two layers have zigzag and bearded edges, respectively. Winding number
calculation shows a topological phase transition between two distinct
non-trivial topological phases when crossing the Dirac points.
Interestingly, we find the stacking configuration of BLG ribbon is important
since they can lead to unexpected edge states without protection from the
chiral symmetry both near the Fermi level in armchair-armchair case and in
the gap within bulk bands that are away from Fermi level in the general
case. The influence of interlayer next nearest neighbor (NNN) interaction
and interlayer bias are also discussed to fit the realistic graphene
materials, which suggest the robust topological features of edge states in
BLG systems.
\end{abstract}

\maketitle


\section{Introduction}

One of the most attractive phenomena in condensed matter physics is the
existence and behavior of edge states, whose wave function is localized at
the system's edge, of two-dimensional (2D) systems. These states are
different from the bulk states in properties and play important roles in
transport, e.g. quantum Hall effect (QHE) and the quantum spin Hall effect
(QSHE) \cite{Kane,Ono,Wu,Bern}.On the other hand, the existence and
properties of zero-energy edge states near the Fermi level (flat bands) are
usually connected with the non-trivial topological phases of the bulk system
by the bulk-edge correspondence \cite{Kane1,Graf,Rud,Bern1,Asb}, which can
be distinguished by the specific symmetry of the system and topological
invariants such as the winding number \cite{Bern1,Asb,Chiu,Gui}.

\begin{figure}[tbp]
\begin{center}
\includegraphics[width=0.48\textwidth]{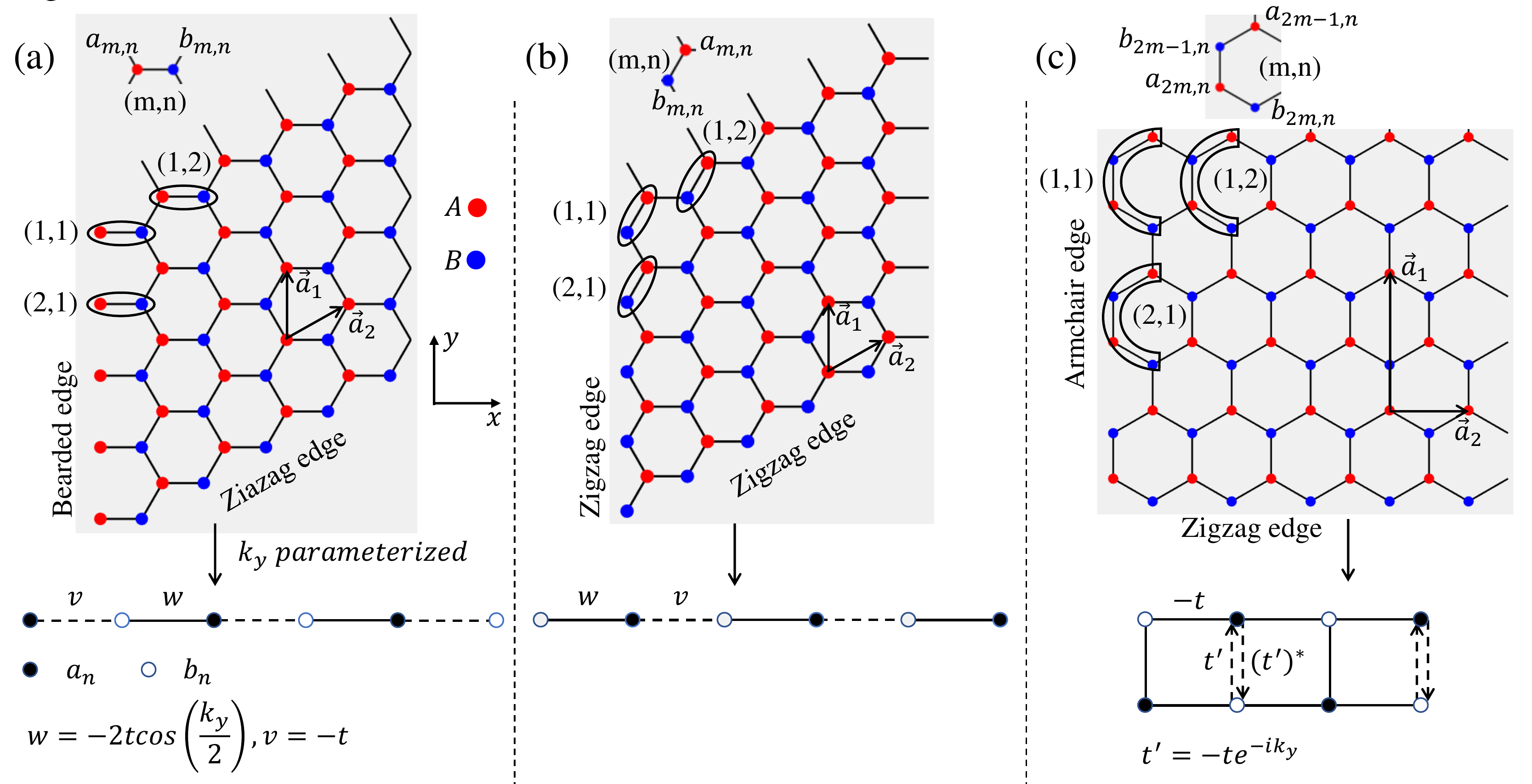}
\end{center}
\caption{(Color online) The schematic illustration of nanoribbons with
different edges and related effective SSH chain parameterized by $k_{y}$.
The primitive vectors are $\vec{a}_{1}$ and $\vec{a}_{2}$. The unit with
translational symmetry in the tight-binding Hamiltonians is emphasized by
the black box in each structures, respectively. The number pairs
(1,1),(1,2),(2,1) in each figure indicate the increasing direction of $m$ and
$n$.}
\label{fig1}
\end{figure}

After the progress in a decade, graphene, or the nanotube and nanoribbon,
has become one of the most active two-dimensional nanomaterial in condensed
matter physics due to excellent electrical and mechanical properties \cite%
{Sai1,Novo1,Mey,Che,Uch,Hua,Zhou,Neto,Dres}. The free standing monolayer
graphene (MLG) is a zero-gap semiconductor where the conduction and valance
band touch each other at the Dirac points \cite%
{Neto,Dres,Wal,Gus,Novo2,Zhang,Kat,Novo3}. It has a trivial bulk topology as
two inequivalent valleys provide opposite topological charges, leading to a
zero Chern number \cite{Yao}. However, edge states still exist in such
graphene systems as non-dispersive bands (flat bands) at the Fermi level
\cite{Neto,Dres,Ryu,Ryc,Yao1,Del,Chiu1}, which are observed by supposing a
semi-infinite system, with quantized wave vectors $k_{y}$ in the infinite
direction \cite{Gui,Ryu,Yao1,Del,Sai}. The existence of these non-dispersive
edge states and related topology in the MLG can be further described by the
bulk-edge correspondence between winding number or Zak phase of
one-dimensional (1D) SSH chain systems and the existence of a localized
state at the chain's edge \cite{Gui,Ryu,Del,Chang-An,SLZhang}, as shown in Fig. \ref{fig1}.

On the other hand, due to the equivalence between the graphene system and the
honeycomb bosonic lattice system, i.e., the 2D magnon system which generally
results from the collinear Ferromagnet after Holstein-Primakoff
transformation \cite{Lie,Kaw,Yok},  edge states similar to those observed in MLG can also
be found in both the related honeycomb bosonic lattices \cite{Pie,Ser}
and even non-honeycomb bosonic lattices \cite{Mao}. There are also both experimental and
theoretical study on other types of edges states in photonic honeycomb lattice using different models. \cite{Plot,Mil}.

In this article, we focus on the existence and topology of edge states in
AB-stacked bilayer graphene (BLG) ribbon \cite{Novo,Mc} by connecting them
with $k_{y}$-parameterized SSH ladder. Earlier works are mainly concentrated
on the existence of edge states of BLG under specific edge conditions \cite%
{Neto1,Mazo} and related equivalent bilayer magnon systems \cite{Sak,Gha},
or the behavior of edge states when various symmetry-breaking terms are
added \cite{Tse,Wei}. Here the AB-stacked BLG ribbon we discuss involves
three conventional edges (zigzag, bearded, and armchair, as shown in Fig.
1), whose bulk 2D system always has zero Chern number. Interlayer next
nearest neighbor (NNN) interaction and interlayer bias are considered in
terms of their influence on the topology of SSH ladder. A detailed
topological classification based on discrete symmetry \cite{Chiu} and
topological invariants calculation for effective 1D bulk Hamiltonian of SSH
ladder $H\left( k_{y},k\right) $ parameterized by $k_{y}$ of AB-stacked BLG
ribbon with various types of edge are performed, as shown in Table \ref%
{Table I}. It shows the zero-energy edge states can only exist when chiral
symmetry is preserved for $H\left( k_{y},k\right) $ and can appear in the
whole $k_{y}$\ region when two layers of BLG ribbon have zigzag and bearded
edge, respectively. On the other hand, straightforward calculation shows
that unexpected edge states can exist in the gap within bulk bands that are
away from the Fermi level. These edge states are unprotected by the chiral
symmetry and are dependent on the specific edge configurations of BLG
ribbon. Interlayer bias is included in our discussion as it explicitly
breaks the chiral symmetry responsible for the existence of zero-energy edge
states. However, edge states still exist after this chiral symmetry breaking
as non-zero energy states.

The rest of paper is organized as follows. We first give a brief review on
the existence of edge states and topology of MLG ribbon in Sec. \ref{Edge
states in the MLG with different edges} as a basis for our discussion of
AB-stacked BLG ribbon. In Sec. \ref{Edge states near the zero energy (Fermi
level) in the AB-stacked BLG}, we turn to the behavior and topology of edge
states in AB-stacked BLG ribbon and their correspondence with SSH ladder $%
H\left( k_{y},k\right) $. Then we discuss the geometrical origin of the edge
states appearing in the gap within bulk bands that are away from the Fermi
level in Sec. \ref{Edge states appeared in the gap away from the zero energy
(Fermi level)}. Finally, we present our conclusions in Sec. \ref{Conclusions
and discussions} as a summary.

\vspace{-0.06cm}

\section{Edge states in the MLG ribbon}

\label{Edge states in the MLG with different edges}

To discuss the existence and topological features of edge states in
AB-stacked BLG ribbon, we first give a brief review of the edge states in
the MLG ribbon. In general, MLG tight-binding Hamiltonian with nearest
neighbor (NN) hopping energy $t$ and on-site potential $U_{i}$ can be
written as \cite{Neto}
\begin{equation}
H=-t\sum_{\left\langle i,j\right\rangle }c_{i}^{\dagger
}c_{j}+\sum_{i}U_{i}c_{i}^{\dagger }c_{i},
\end{equation}%
where $\sum_{\left\langle i,j\right\rangle }$ sums over only NN pairs. The
lattice primitive vectors are $\overrightarrow{a}_{1}$ and $\overrightarrow{a%
}_{2}$, which are shown in Fig. \ref{fig1}. As examples and without loss of
generality, we mainly consider nanoribbons with three different types of
edges: zigzag, bearded, and armchair in $y$ direction and enforce the
periodic boundary condition (PBC) along this direction to see the edge
states, as shown in Fig. \ref{fig1}(a), (b), and (c), respectively.

\subsection{MLG with bearded (zigzag)\ edges}

The tight-binding Hamiltonian of a MLG with a bearded edge in $y$ direction,
as shown in Fig. \ref{fig1}(a), can be expressed as%
\begin{eqnarray}
H_{\mathrm{bea}} &=&-t\sum_{m=1}^{M}\{\sum_{n=1}^{N}[a_{m,n}^{\dagger
}b_{m,n}+b_{m,n}^{\dagger }\left( a_{m,n+1}+a_{m+1,n+1}\right) ]  \notag \\
&&-b_{m,N}^{\dagger }\left( a_{m,1}+a_{m+1,1}\right) +\mathrm{H.c.}\},
\end{eqnarray}%
where $a_{m,n}\left( a_{m,n}^{\dagger }\right) $ annihilates (creates) an
electron on site $\left( m,n\right) $ on sublattice $A$ (an equivalent
definition is used for sublattice $B$). The system is assumed infinite along
m direction and finite along n direction. The minus term in the curly
brackets gives the open boundary condition (OBC) in the finite direction.
The Fourier transformation along the infinite direction is%
\begin{eqnarray}
f_{m,n} &=&\frac{1}{\sqrt{M}}\sum_{k_{y}}e^{ik_{y}m}f_{k_{y},n,},f=a,b, \\
k_{y} &=&\frac{2\pi \left( m-M/2\right) }{M},m=0,1,2,\ldots ,M-1,  \notag
\end{eqnarray}%
and leads to%
\begin{eqnarray}
H_{\mathrm{bea}}\left( k_{y}\right) &=&-t\{\sum_{n=1}^{N-1}\left[
a_{k_{y},n}^{\dagger }b_{k_{y},n}+\left( 1+e^{ik_{y}}\right)
b_{k_{y},n}^{\dagger }a_{k_{y},n+1}\right]  \notag \\
&&+a_{k_{y},N}^{\dagger }b_{k_{y},N}+\mathrm{H.c.}\},
\end{eqnarray}%
which is equivalent to an effective SSH chain parameterized by $k_{y}$ as
below. Notice that we have made a redefinition of basis by a phase such that
the hopping becomes real and it is easier to make association with the
original SSH chain model. Not doing such a redefinition would leave the
coupling complex, but all results in this paper are not affected.
\begin{eqnarray}
H_{\mathrm{bea}}\left( k_{y}\right) &\simeq &\sum_{n=1}^{N-1}\left(
va_{n}^{\dagger }b_{n}+wb_{n}^{\dagger }a_{n+1}\right) +va_{N}^{\dagger
}b_{N}+\mathrm{H.c.},  \notag \\
w &=&-2t\cos \frac{k_{y}}{2},v=-t.
\end{eqnarray}

The bulk Hamiltonian of this equivalent chain is%
\begin{eqnarray}
H_{\mathrm{bea}}\left( k_{y},k\right) &=&\eta ^{\dagger }h_{\mathrm{b}%
}\left( k_{y},k\right) \eta ,\eta =\left( a_{k},b_{k}\right) ^{T},
\label{hb} \\
h_{\mathrm{b}}\left( k_{y},k\right) &=&\left[
\begin{array}{cc}
0 & v+we^{-ik} \\
v+we^{ik} & 0%
\end{array}%
\right] ,  \notag
\end{eqnarray}%
which belongs to the non-trivial topological class $\mathcal{BDI}$ (see
Table. \ref{Table I} for details).

\begin{figure}[tbp]
\begin{center}
\includegraphics[width=0.48\textwidth]{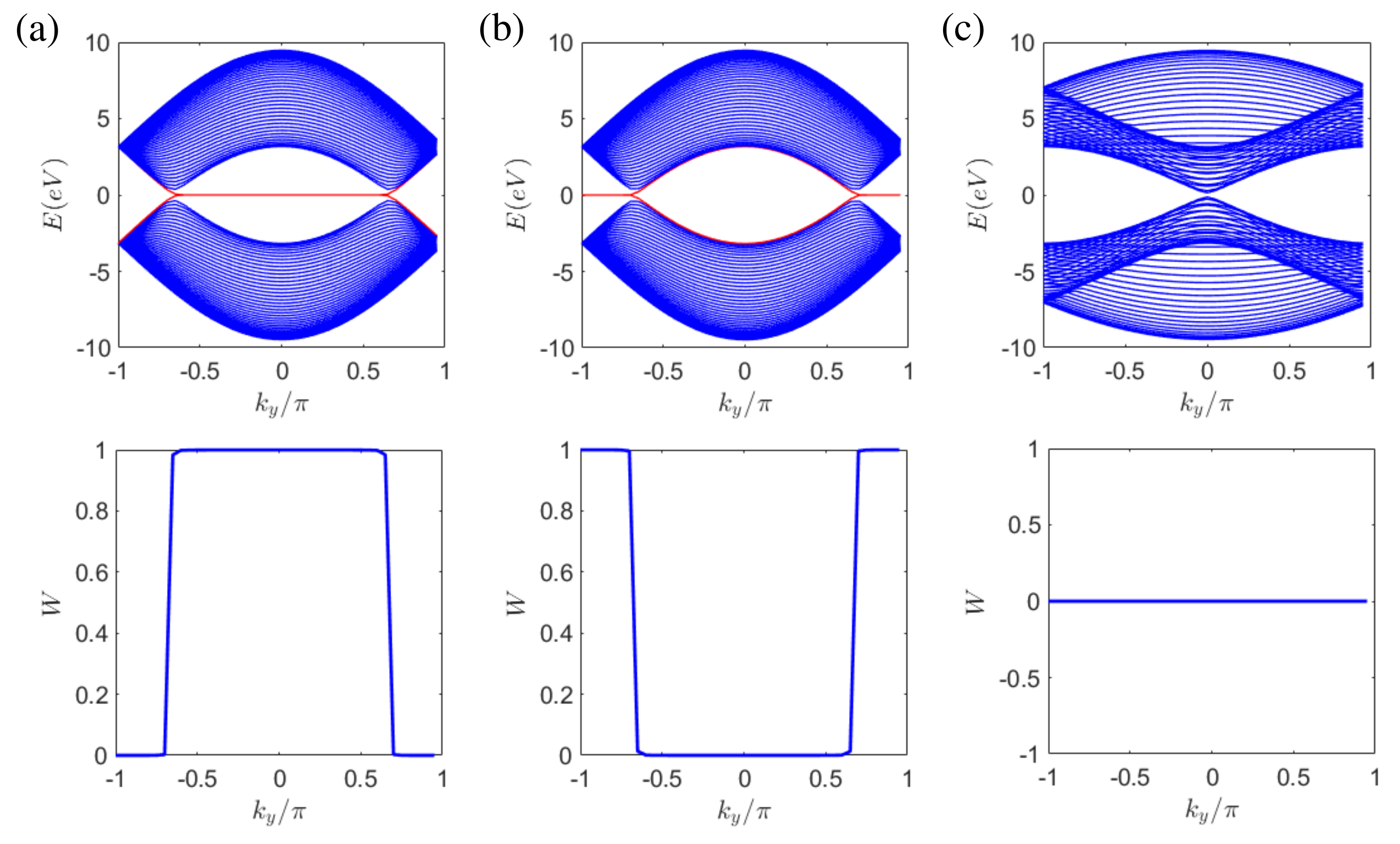}
\end{center}
\caption{(Color online) The band structure of MLG ribbon and related winding
number $W(k_{y})$ for effective $k_{y}$-parameterized SSH chain. (a) for the
MLG with bearded edges, (b) for the MLG with zigzag edges, and (c) for the
MLG with armchair edges, respectively. Here the nearest neighbor interaction
is chosen as $t=3.16\mathrm{eV}$ according to Ref. \protect\cite{Kuz}. The
red solid lines represent the edge states. }
\label{fig2}
\end{figure}

We plot the band structure of bearded-edge graphene nanoribbon and winding
number for the bulk Hamiltonian $H_{\mathrm{bea}}\left( k_{y},k\right) $ (%
\ref{hb}) as a function of parameter $k_{y}$ in Fig. \ref{fig2}(a),
respectively. The winding number we used in this paper is defined as \cite%
{Chiu}%
\begin{equation}
W=-\frac{i}{4\pi }\int_{BZ}\mathrm{d}k\mathrm{Tr}\left( SQ^{-1}\partial
_{k}Q\right) ,
\end{equation}%
with%
\begin{equation}
Q\left( k\right) =\mathit{I}_{N}-2\mathcal{P}\left( k\right) ,\mathcal{P}%
\left( k\right) =\sum_{\alpha <0}\left\vert u_{\alpha }\right\rangle
\left\langle u_{\alpha }\right\vert ,
\end{equation}%
$\alpha <0$ refers to the occupied bands (eigenenstates of $H\left(
k_{y},k\right) $ below the Fermi level). $Q^{-1}\left( k\right) =Q\left(
k\right) $, as $Q^{2}=\mathit{I}_{N}$. Winding number describes the
topological properties near the zero energy (Fermi level) of 1D bulk
Hamiltonians $H\left( k_{y},k\right) $ with the chiral symmetry operator $S$%
\cite{Chiu} . Here we would like to stress again one should not confuse $%
k_{y}$ with $k$, since $k_{y}$ appearing in the bulk Hamiltonian $H\left(
k_{y},k\right) $ is a parameter of the system and $k$ is the wave vector of
the effective SSH chain when its length is taken to be infinite. The
integration appearing in the definition of $W$ is over $k$, with $W$ being a
function of $k_{y}$. A concrete example with some details omitted here is given in the Appendix to make clearer the origin of $k_y$ and $k$. The zero-energy edge states (flat bands) exist in the
restricted region $k_{y}\in \left[ -\pi ,-\frac{2}{3}\pi \right] \cup \left[
\frac{2}{3}\pi ,\pi \right] $ and correspond $W=1$, which agree with
previous literatures \cite{Yao,Del}. The effective 1D Hamiltonian $H_{%
\mathrm{zig}}\left( k_{y}\right) $ for a MLG with a zigzag edge can be
obtained by switching $w$ and $v$, $a_{n\text{ }}$and $b_{n}$ in $H_{\mathrm{%
bea}}\left( k_{y}\right) $ as zigzag edge is related with bearded edge
through an exchange of basis and coupling (see Fig. \ref{fig1}(a) and (b)).
In terms of their corresponding SSH chain, zigzag nanoribbon and bearded
nanoribbon differ from each other by a switch between intercell coupling and
intracell coupling of the chain. It can be observed from Fig. \ref{fig2}(a)
and (b) that zero-energy edge states (flat bands) of $H_{\mathrm{zig}}\left(
k_{y}\right) $ and non-zero winding number of $H_{\mathrm{zig}}\left(
k_{y},k\right) $ appear in complementary region of $k_{y}$ to the one with
bearded edges.

\vspace{-0.06cm}

\subsection{MLG with armchair edges}

The tight-binding Hamiltonian of a MLG ribbon with an armchair edge is
different from the previous case, which is%
\begin{eqnarray}
H_{\mathrm{arm}} &=&-t\sum_{n=1}^{N}[\sum_{m=1}^{M/2}(a_{2m,n}^{\dagger
}b_{2m,n}+b_{2m,n}^{\dagger }a_{2m,n+1} \\
&&+a_{2m-1,n}^{\dagger }b_{2m-1,n}+b_{2m-1,n+1}^{\dagger }a_{2m-1,n})  \notag
\\
&&+\sum_{m=1}^{M}a_{m+1,n}^{\dagger }b_{m,n}]  \notag \\
&&+\sum_{m=1}^{M/2}t\left( b_{2m,N}^{\dagger }a_{2m,1}+b_{2m-1,1}^{\dagger
}a_{2m-1,N}\right) +\mathrm{H.c.},  \notag
\end{eqnarray}%
where $M/2\in
\mathbb{N}
$. Unit with translational symmetry of this ribbon in Fig. \ref{fig1}(c) is
constructed as $\left( m,n\right) \simeq \left(
b_{2m-1,n},a_{2m-1,n},a_{2m,n},b_{2m,n}\right) $. The effective Hamiltonian
parameterized by $k_{y}$ for this ribbon is no longer a single SSH chain but
two coupled uniform chains as shown in Fig. \ref{fig1}(c). The coupled
chains have the bulk Hamiltonian:%
\begin{eqnarray}
H_{\mathrm{arm}}\left( k_{y},k\right) &=&\eta ^{\dagger }h_{\mathrm{a}%
}\left( k_{y},k\right) \eta ,\eta =\left(
b_{k}^{1},a_{k}^{1},a_{k}^{2},b_{k}^{2}\right) ^{T},  \notag \\
h_{\mathrm{a}}\left( k_{y},k\right) &=&-t\left[
\begin{array}{cc}
h\left( k\right) & D\left( k_{y}\right) \\
D^{\ast }\left( k_{y}\right) & h\left( k\right)%
\end{array}%
\right] ,  \notag \\
h\left( k\right) &=&\left[
\begin{array}{cc}
0 & 1+e^{-ik} \\
1+e^{ik} & 0%
\end{array}%
\right] ,  \notag \\
D\left( k_{y}\right) &=&\left[
\begin{array}{cc}
1 & 0 \\
0 & e^{-ik_{y}}%
\end{array}%
\right] ,  \label{ha}
\end{eqnarray}%
where the superscript $1/2$ distinguishes even and odd since there are two
sets of A/B in each unit of armchair MLG shown in Fig. \ref{fig1}(c):%
\begin{equation}
\begin{split}
f_{k}^{1}& =\frac{1}{\sqrt{N}}\frac{1}{\sqrt{M/2}}\sum_{j=1}^{M/2}%
\sum_{n=1}^{N}e^{-ikn}e^{-ik_{y}j}f_{2j-1,n} \\
f_{k}^{2}& =\frac{1}{\sqrt{N}}\frac{1}{\sqrt{M/2}}\sum_{j=1}^{M/2}%
\sum_{n=1}^{N}e^{-ikn}e^{-ik_{y}j}f_{2j,n}
\end{split}%
\end{equation}%
$f=a,b$ as shown in Fig. \ref{fig1}. Notice, again, $k$ is the wave vector
of the coupled chains while $k_{y}$ is a parameter of its coupling. This
bulk Hamiltonian of the coupled chains belongs to the non-trivial
topological class $\mathcal{AIII}$ (see Table \ref{Table I} for detail).
Here we would like to point out that although it belongs to the non-trivial
topological class, the winding number is zero in the whole region of $k_{y}$
and there are no edge states, as shown in Fig. \ref{fig2}(c), which means
this is a trivial case.

\begin{figure}[tbp]
\begin{center}
\includegraphics[width=0.48\textwidth]{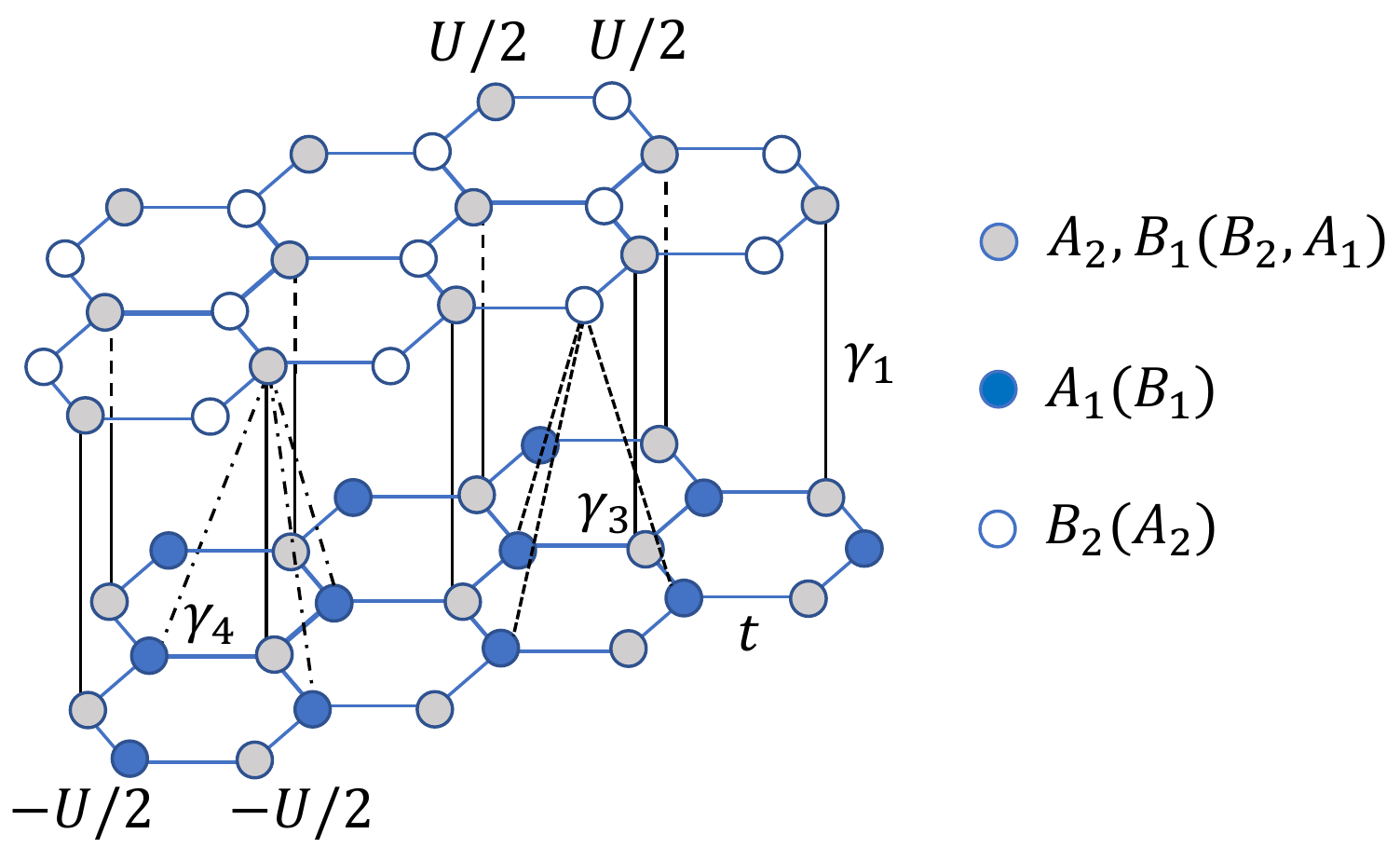}
\end{center}
\caption{(Color online) The atomic structure of AB-stacked BLG in side view.}
\label{fig3}
\end{figure}
\vspace{-0.06cm}

\begin{table*}[tbph]
\caption{Topological classification for different effective 1D bulk
Hamiltonians $H(k_{y},k)$ of SSH chain/ladder. The related winding number $W$
and number of zero-energy edge states $N_{ES}$ are also shown. The
topological classification is based on the the presence or absence ($0$) of
time-reversal ($T$), particle-hole ($C$), and chiral ($S$) symmetries
\protect\cite{Chiu}, where all three symmetry operators are unitary, i.e., $%
O^{\dagger }O=1,O=T,C,S$ . They satisfy $T^{\dagger }H^{\ast
}(k_{y},k)T=H(k_{y},-k)$, $C^{\dagger }H^{\ast }(k_{y},k)C=-H(k_{y},-k)$, $%
S^{\dagger }H(k_{y},k)S=-H(k_{y},k)$, respectively. $\pm $ in $T$ and $C$
comes from $T^{\ast }T=\pm 1$ and $C^{\ast }C=\pm 1$. $\mathit{I}_{N}$ is $%
N\times N$ identity matrix. $\protect\sigma _{\protect\alpha =x,y,z}$
represent Pauli matrices. \textquotedblleft -\textquotedblright\ means that
there is no well-defined winding number since the chiral symmetry is broken
\protect\cite{Chiu}. Notice that when operator is written in direct product
form, they can be understood as acting on different degrees of freedom
(DOF). For example, $S=\mathit{I}_{2}\otimes \protect\sigma _{z}$ for $H_{%
\mathrm{bea-bea}}$, where $I_{2}$ acts on layer DOF and $\protect\sigma _{z}$
acts on sublattice DOF.}
\label{Table I}
\begin{center}
\setlength{\arrayrulewidth}{0.5mm} 
\renewcommand\tabcolsep{5.5pt} 
\begin{tabular}{ccccccc}
\hline\hline
&  &  &  &  &  &  \\[-1ex]
Effective 1D bulk Hamiltonian & $T$ & $C$ & $S$ & Class & $W$ & $N_{ES}$ \\%
[0.5ex] \hline
&  &  &  &  &  &  \\[-1ex]
$H_{\mathrm{bea/zig}}(k_{y},k)$ (Eq. \ref{hb}) & $\mathit{I}_{2}(+)$ & $%
\sigma _{z}(+)$ & $\sigma _{z}$ & $\mathcal{BDI}$ & $1,0$ (Fig. \ref{fig2})
& $2$ (Fig. \ref{fig2}) \\[0.5ex]
$H_{\mathrm{arm}}(k_{y},k)$ (Eq. \ref{ha}) & $0$ & $0$ & $\sigma _{z}\otimes
\sigma _{z}$ & $\mathcal{AIII}$ & $0$ (Fig. \ref{fig2}) & $0$ (Fig. \ref%
{fig2}) \\[0.5ex]
$H_{\mathrm{arm-arm}}^{\uparrow }(k_{y},k),U=0$ (Eq. \ref{haa}) & $0$ & $0$
& $\mathit{I}_{2}\otimes \sigma _{z}\otimes \sigma _{z}$ & $\mathcal{AIII}$
& $0$ & $0$ (Fig. \ref{fig4}) \\[0.5ex]
$H_{\mathrm{arm-arm}}^{\uparrow }(k_{y},k),U\neq 0$ (Eq. \ref{haa}) & $0$ & $%
0$ & $0$ & $\mathcal{A}$ & - & $0$ (Fig. \ref{fig4}) \\[0.5ex]
$H_{\mathrm{arm-arm}}^{\swarrow }(k_{y},k),U=0$ (Eq. \ref{haa}) & $0$ & $0$
& $\mathit{I}_{2}\otimes \sigma _{z}\otimes \sigma _{z}$ & $\mathcal{AIII}$
& $0$ & $0$ (Fig. \ref{fig5}) \\[0.5ex]
$H_{\mathrm{arm-arm}}^{\swarrow }(k_{y},k),U\neq 0$ (Eq. \ref{haa}) & $0$ & $%
0$ & $0$ & $\mathcal{A}$ & - & $4$ (Fig. \ref{fig5}) \\[0.5ex]
$H_{\mathrm{bea-bea/zig-zig}}^{\leftarrow (\swarrow )}(k_{y},k),U=0$ (Eq. %
\ref{hbb}) & $\mathit{I}_{4}(+)$ & $\mathit{I}_{2}\otimes \sigma _{z}(+)$ & $%
\mathit{I}_{2}\otimes \sigma _{z}$ & $\mathcal{BDI}$ & $2,0$ (Fig. \ref{fig7}%
) & $4$ (Fig. \ref{fig7}) \\[0.5ex]
$H_{\mathrm{bea-bea/zig-zig}}^{\leftarrow (\swarrow )}(k_{y},k),U\neq 0$
(Eq. \ref{hbb}) & $\mathit{I}_{4}(+)$ & $0$ & $0$ & $\mathcal{AI}$ & - & $4$
(Fig. \ref{fig7}) \\[0.5ex]
$H_{\mathrm{bea-zig}}^{\swarrow (\nearrow )}(k_{y},k),U=0$ (Eq. \ref{hbz}) &
$\mathit{I}_{4}(+)$ & $\sigma _{z}\otimes \sigma _{z}(+)$ & $\sigma
_{z}\otimes \sigma _{z}$ & $\mathcal{BDI}$ & $-1,1$ (Fig. \ref{fig7}) & $2$
(Fig. \ref{fig7}) \\[0.5ex]
$H_{\mathrm{bea-zig}}^{\swarrow (\nearrow )}(k_{y},k),U\neq 0$ (Eq. \ref{hbz}%
) & $\mathit{I}_{4}(+)$ & $0$ & $0$ & $\mathcal{AI}$ & - & $2$ (Fig. \ref%
{fig7}) \\[0.5ex] \hline
\end{tabular}%
\end{center}
\end{table*}

\section{Edge states near the zero energy (Fermi level) in the AB-stacked BLG%
}

\label{Edge states near the zero energy (Fermi level) in the AB-stacked BLG}

Based on the three different graphene nanoribbons discussed in last section,
there are four types of AB-stacked BLG ribbon, i.e., arm-arm (both layers are armchair edges), zig-zig/bea-bea (both layers are
zigzag or bearded edges), bea-zig (one layer is bearded edges and the other is zigzag edges), as listed in Table. \ref{Table I}.
They can be summarized by the tight-binding Hamiltonian \cite{Neto,Mc,Mc1}%
\begin{equation}
H=\sum_{l}H_{\mathrm{edge}}^{l}+H_{\mathrm{int}}+H_{\mathrm{on-site}},
\end{equation}%
where $l=1,2$ are labels of the bottom and top layers respectively. $%
H_{\mathrm{on-site}}$ refers to the on-site energy of carbon atoms such as
interlayer bias $U$ as shown in Fig. \ref{fig3}. It includes contribution from
both layers. $H_{\mathrm{int}}$ describes the van der Waals interaction
between two layers \cite{Neto,Mc,Mc1}. The meanings of various interlayer
couplings $\gamma _{i} $ with $i=1,3,4$ are indicated in Fig. \ref{fig3}. Here, we take $t=3.16%
\mathrm{eV}$, $\gamma _{1}=0.381\mathrm{eV}$ as typical experimental values
for AB-stacked BLGs \cite{Kuz}, and $\gamma _{4}=0$ throughout this work
since it is pretty small compared with others in realistic systems \cite%
{Neto,Mc1,Kuz}. We choose $\gamma _{3}=0.38\mathrm{eV}\approx \gamma _{1}$
when considering the non-zero $\gamma _{3}$, which is close to most of the
experimental observations \cite{Neto,Mc1,Kuz}$.$

\begin{figure}[tbp]
\begin{center}
\includegraphics[width=0.48\textwidth]{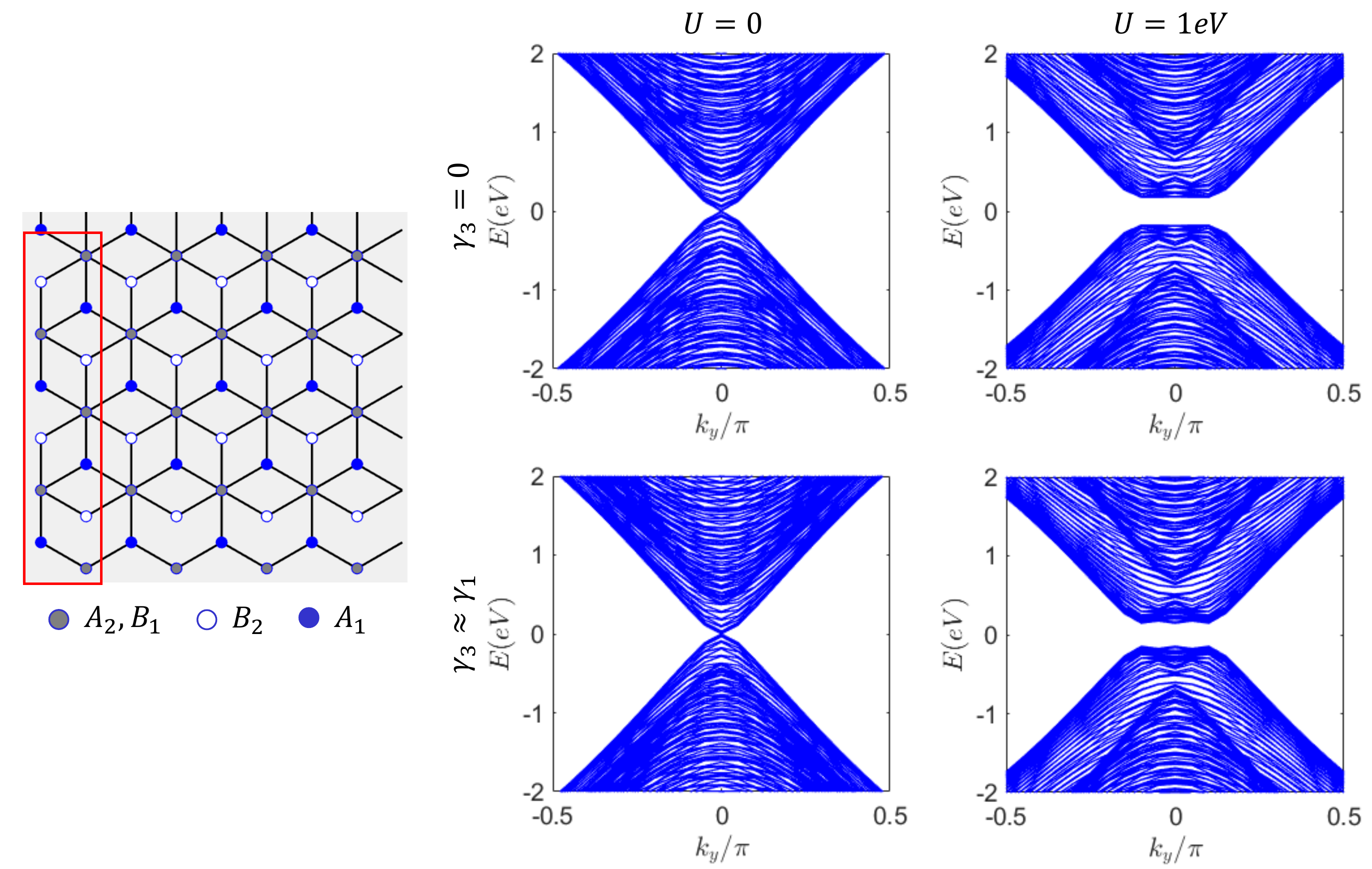}
\end{center}
\caption{(Color online) Left panel: The atomic structure of an
armchair-armchair AB-stacked BLG ribbon in top view, corresponding to 1D
effective Hamiltonian $H_{\mathrm{arm-arm}}^{\uparrow }(k_{y},k)$ (Eq.
\protect\ref{haa}). The edge configuration is emphasized by the red box.
Right panel: Related band structure with different interlayer bias $U$ and
NNN interaction $\protect\gamma _{3}$.}
\label{fig4}
\end{figure}
\vspace{-0.06cm}

\begin{figure}[tbp]
\begin{center}
\includegraphics[width=0.48\textwidth]{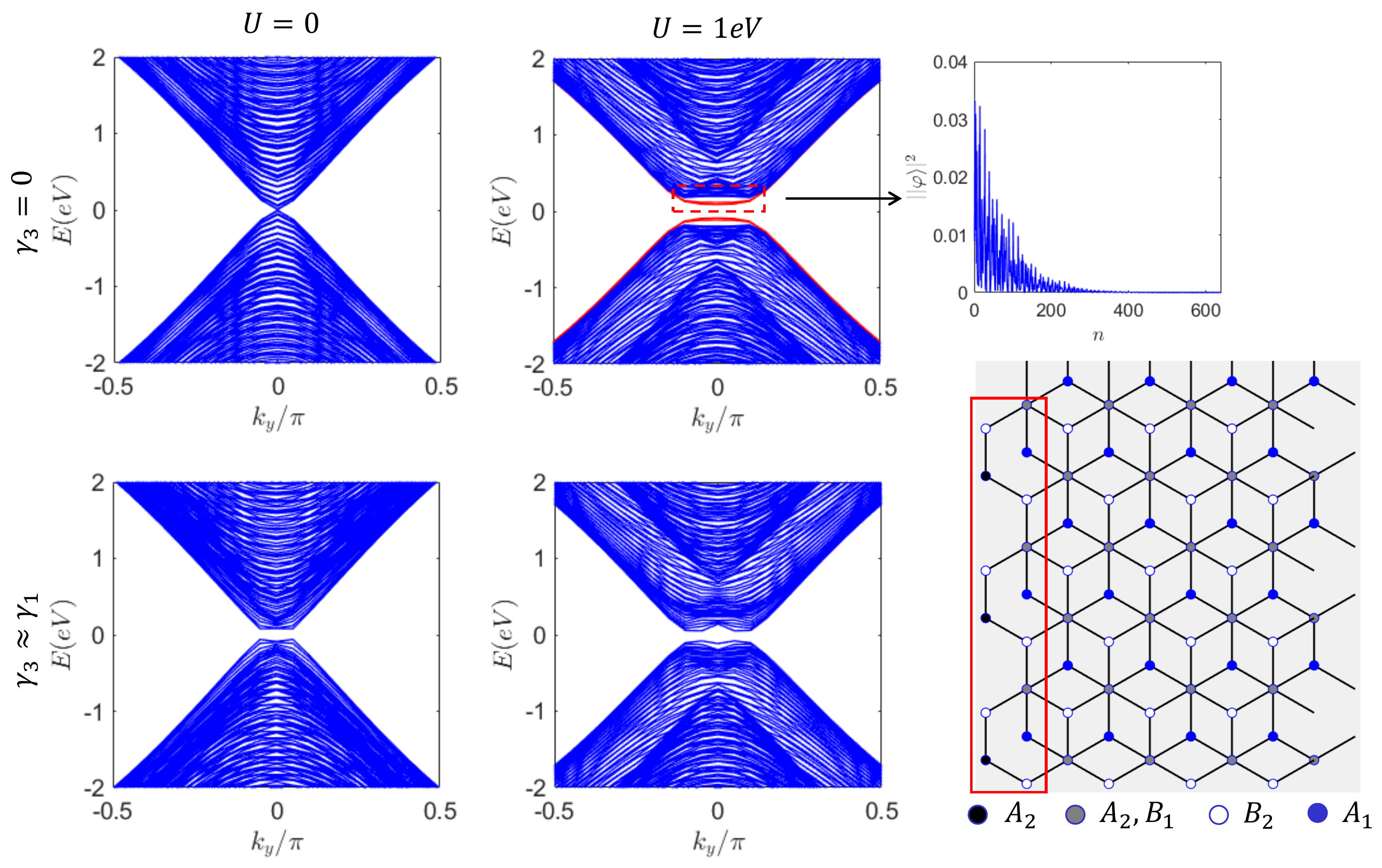}
\end{center}
\caption{(Color online) Right panel: The atomic structure of an
armchair-armchair AB-stacked BLG ribbon in top view, corresponding to 1D
effective Hamiltonian $H_{\mathrm{arm-arm}}^{\swarrow }(k_{y},k)$ (Eq.
\protect\ref{haa}). The edge configuration is emphasized by the red box,
which is different from the one in Fig. \protect\ref{fig4}. Left panel:
Related band structure with different interlayer bias $U$ and NNN
interaction $\protect\gamma _{3}$. The red solid lines are the edge states.
There are four of them. The wave function distribution in real space of a
typical edge state in red dashed box is shown on the right side. Vertical
axis is wave function's amplitude, and the horizontal axis is the site
index, which increases along the finite direction of the ribbon. }
\label{fig5}
\end{figure}
\vspace{-0.06cm}

\subsection{AB-stacked BLG with armchair-armchair edges}

Next we will discuss these four types of AB-stacked BLG structures in
detail. To form an AB-stacked BLG ribbon, armchair MLG can only be stacked
with the other armchair MLG, but not with zigzag/bearded MLG (see Fig. \ref%
{fig1}). Thus, we first consider the Hamiltonian%
\begin{equation}
H_{\mathrm{arm-arm}}=\sum_{l}H_{\mathrm{arm}}^{l}+H_{\mathrm{int}}+H_{%
\mathrm{on-site}},
\end{equation}%
where both layers have armchair edge. We further consider two different edge
configurations as shown in left panel of Fig. \ref{fig4} and \ref{fig5},
corresponding to different $H_{\mathrm{int}}$. The band structure of $H_{%
\mathrm{arm-arm}}^{\uparrow \left( \swarrow \right) }\left( k_{y}\right) $
parameterized by $k_{y}$ is shown in right panel of Fig. \ref{fig4} and \ref%
{fig5} respectively, where the label $\uparrow \left( \swarrow \right) $
represents the different edge configuration of two setting. It denotes the
direction to translate top layer if we effectively consider the AB-stacked
nanoribbon is formed from the relative translation between two layers of
AA-stacked nanoribbon. Similar calculation to that of last section shows the
related $k_y$-parameterized bulk Hamiltonian is%
\begin{eqnarray}
H_{\mathrm{arm-arm}}^{\uparrow \left( \swarrow \right) }\left(
k_{y},k\right) &=&\eta ^{\dagger }h_{\mathrm{aa}}^{\uparrow \left( \swarrow
\right) }\left( k_{y},k\right) \eta ,\eta =\left( \xi _{1},\xi _{2}\right)
^{T},  \label{haa} \\
\xi _{l} &=&\left( b_{l,k}^{1},a_{l,k}^{1},a_{l,k}^{2},b_{l,k}^{2}\right) ,
\notag \\
h_{\mathrm{aa}}^{\uparrow \left( \swarrow \right) }\left( k_{y},k\right) &=&%
\left[
\begin{array}{cc}
h_{\mathrm{a}}^{1}\left( k_{y},k\right) & -H_{\mathrm{int}}^{\uparrow \left(
\swarrow \right) }\left( k_{y},k\right) \\
-\left[ H_{\mathrm{int}}^{\uparrow \left( \swarrow \right) }\left(
k_{y},k\right) \right] ^{\dagger } & h_{\mathrm{a}}^{2}\left( k_{y},k\right)%
\end{array}%
\right]  \notag \\
h_{\mathrm{a}}^{l}\left( k_{y},k\right) &=&\frac{(-1)^{l}U}{2}\mathit{I}%
_{4}+h_{\mathrm{a}}\left( k_{y},k\right) ,f_{k}=1+e^{ik}  \notag \\
H_{\mathrm{int}}^{\uparrow }\left( k_{y},k\right) &=&\left[
\begin{array}{cccc}
0 & 0 & \gamma _{1} & 0 \\
\gamma _{3}f_{k} & 0 & 0 & \gamma _{3} \\
\gamma _{3}e^{ik_{y}} & 0 & 0 & \gamma _{3}f_{k}^{\ast } \\
0 & \gamma _{1}e^{ik_{y}} & 0 & 0%
\end{array}%
\right] ,  \notag \\
H_{\mathrm{int}}^{\swarrow }\left( k_{y},k\right) &=&\left[
\begin{array}{cccc}
0 & \gamma _{1} & 0 & 0 \\
\gamma _{3}e^{ik} & 0 & 0 & \gamma _{3}f_{k}e^{-ik_{y}} \\
\gamma _{3}f_{k} & 0 & 0 & \gamma _{3} \\
0 & 0 & \gamma _{1}e^{ik} & 0%
\end{array}%
\right] ,  \notag
\end{eqnarray}%
which belongs to the non-trivial topological class $\mathcal{AIII}$ when
there is no interlayer bias. When $U\neq 0$, it belongs to the trivial
topological class $\mathcal{A}$ (Table. \ref{Table I}).

For the BLG ribbon with the edge configuration shown in Fig. \ref{fig4},
there is no edge state even if non-zero $U$ and $\gamma _{3}$ are
considered. For the other edge configuration as shown in Fig. \ref{fig5},
the edge states which are gapped and flat appear when non-zero interlayer
bias is added. Here we would like to point out that these edge states are
not topological for three reasons: (i) They cannot be described by the
bulk-edge correspondence we used before since their energy is not at the
Fermi level. (ii) They are not robust since they disappear when including $%
\gamma _{3}$, the quantity that preserves original chiral symmetry of the
system and do not influence the existence of edge states in the case of
bilayer bearded-bearded (zigzag-zigzag) ribbon (discussed below). (iii) Most
importantly, they are not formed between two Dirac points with different
topological charges \cite{Yao,Yao1,Del,Man} in the band structure. But this
fact is still interesting since it indicates the existence of edge states
can be determined by interlayer bias.

\begin{figure*}[tbp]
\begin{center}
\includegraphics[width=0.48\textwidth]{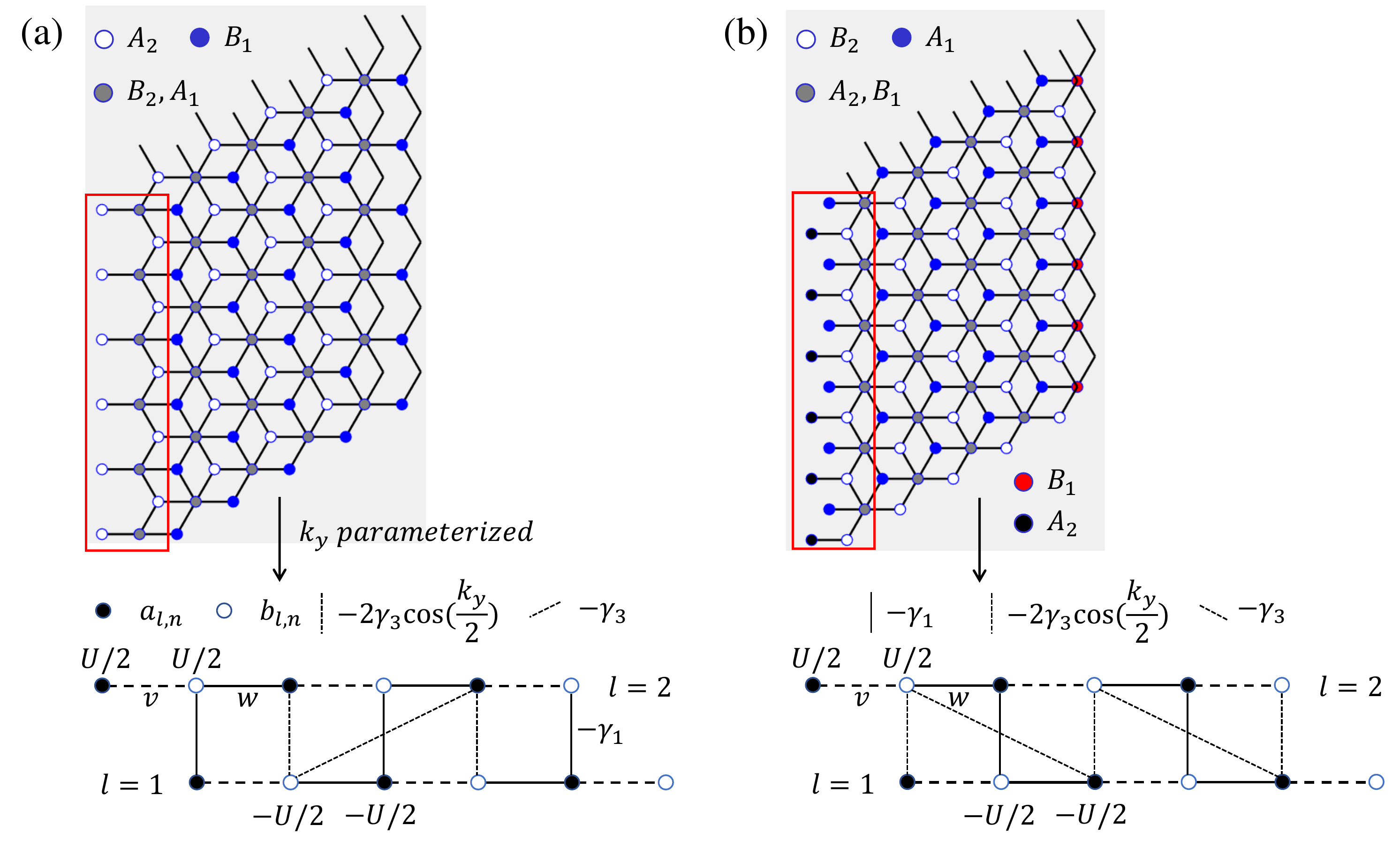} %
\includegraphics[width=0.48\textwidth]{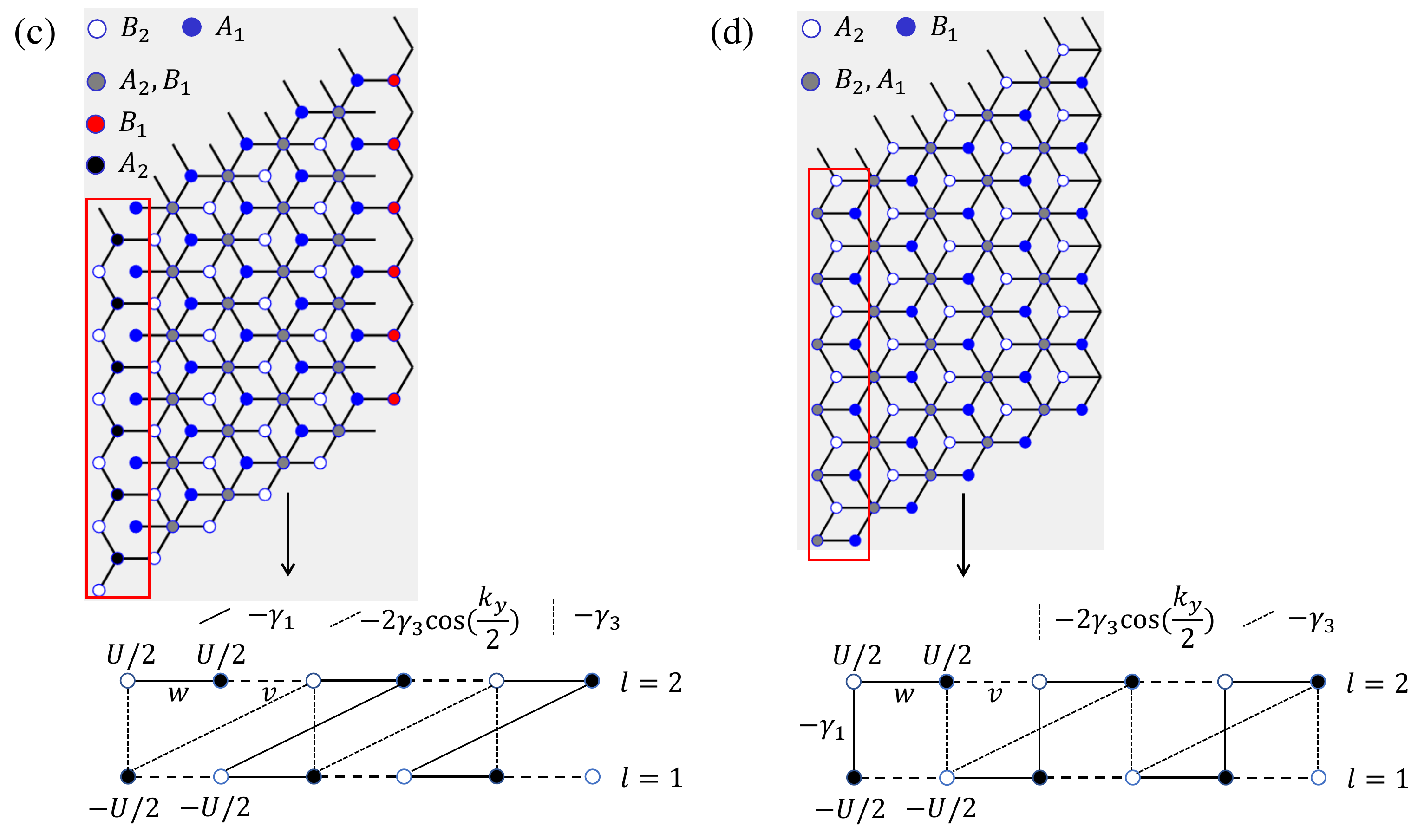}
\end{center}
\caption{(Color online) The atomic structure of bearded-bearded ((a) and
(b)) and bearded-zigzag ((c) and (d)) AB-stacked BLG ribbon in top view. The
specific edge configuration is emphasized by the red box. The related $k_{y}$%
-parameterized SSH ladders are shown at the bottom, where (a)-(b)
corresponds to $H_{\mathrm{ber-ber}}^{\leftarrow (\swarrow )}(k_{y},k)$ (%
\protect\ref{hbb}) and (c)-(d) corresponds to $H_{\mathrm{ber-zig}%
}^{\swarrow (\nearrow )}(k_{y},k)$ (\protect\ref{hbz}), respectively. $%
w=-2tcos(k_{y}/2)$ and $v=-t$ are the same as used in Fig. \protect\ref{fig1}%
.}
\label{fig6}
\end{figure*}
\vspace{-0.06cm}

\subsection{AB-stacked BLG with bearded-bearded (zigzag-zigzag) edges}

The Hamiltonian that both layers have a bearded edge is analogous to the one
with a zigzag edge. We take the one with the bearded edge as an example,
whose Hamiltonian can be expressed as%
\begin{equation}
H_{\mathrm{bea-bea}}=\sum_{l}H_{\mathrm{bea}}^{l}+H_{\mathrm{int}}+H_{%
\mathrm{on-site}}.
\end{equation}%
There are two different edge configurations, corresponding to different
forms of $H_{\mathrm{int}}$. Here these two types of Hamiltonians are
denoted as $H_{\mathrm{bea-bea}}^{\longleftarrow \left( \swarrow \right)
}\left( k_{y}\right) $ . Corresponding lattice structures are shown in Fig. %
\ref{fig6}(a) and (b), respectively. Meanwhile, $H_{\mathrm{zig-zig}%
}^{\longleftarrow \left( \swarrow \right) }\left( k_{y}\right) $ are
obtained by the substitution: $w\longleftrightarrow v$, $a_{l,n}%
\longleftrightarrow b_{l,n}$ in Fig. \ref{fig6}(a) and (b). The bulk
Hamiltonian of the $k_{y}$-parameterized SSH ladder $H_{\mathrm{bea-bea}%
}^{\longleftarrow \left( \swarrow \right) }\left( k_{y}\right) $ can be
expressed as%
\begin{eqnarray}
H_{\mathrm{bea-bea}}^{\longleftarrow \left( \swarrow \right) }\left(
k_{y},k\right) &=&\eta ^{\dagger }h_{\mathrm{bb}}^{\longleftarrow \left(
\swarrow \right) }\left( k_{y},k\right) \eta ,  \label{hbb} \\
\eta &=&\left( a_{1,k},b_{1,k},a_{2,k},b_{2,k}\right) ^{T},  \notag \\
h_{\mathrm{bb}}^{\longleftarrow \left( \swarrow \right) }\left(
k_{y},k\right) &=&\left[
\begin{array}{cc}
h_{\mathrm{b}}^{1}\left( k_{y},k\right) & H_{\mathrm{int}}^{\longleftarrow
\left( \swarrow \right) }\left( k_{y},k\right) \\
\left[ H_{\mathrm{int}}^{\longleftarrow \left( \swarrow \right) }\left(
k_{y},k\right) \right] ^{\dagger } & h_{\mathrm{b}}^{2}\left( k_{y},k\right)%
\end{array}%
\right] ,  \notag
\end{eqnarray}%
with%
\begin{equation}
h_{\mathrm{b}}^{l}\left( k_{y},k\right) =h_{\mathrm{b}}\left( k_{y},k\right)
+\frac{(-1)^{l}}{2}U\mathit{I}_{2},l=1,2,
\end{equation}%
\begin{equation*}
H_{\mathrm{int}}^{\longleftarrow }\left( k_{y},k\right) =-\left[
\begin{array}{cc}
0 & \gamma _{1} \\
\gamma _{3}e^{ik}\left( 2\cos \frac{k_{y}}{2}+e^{ik}\right) & 0%
\end{array}%
\right] ,
\end{equation*}%
\begin{equation*}
H_{\mathrm{int}}^{\swarrow }\left( k_{y},k\right) =-\left[
\begin{array}{cc}
0 & \gamma _{3}\left( 2\cos \frac{k_{y}}{2}+e^{-ik}\right) \\
\gamma _{1}e^{ik} & 0%
\end{array}%
\right] .
\end{equation*}%
which belongs to non-trivial class $\mathcal{BDI}$ only when $U=0$,
otherwise it belongs to trivial class $\mathcal{AI}$ (Table. \ref{Table I}).

\begin{figure*}[tbp]
\begin{center}
\includegraphics[width=0.7\textwidth]{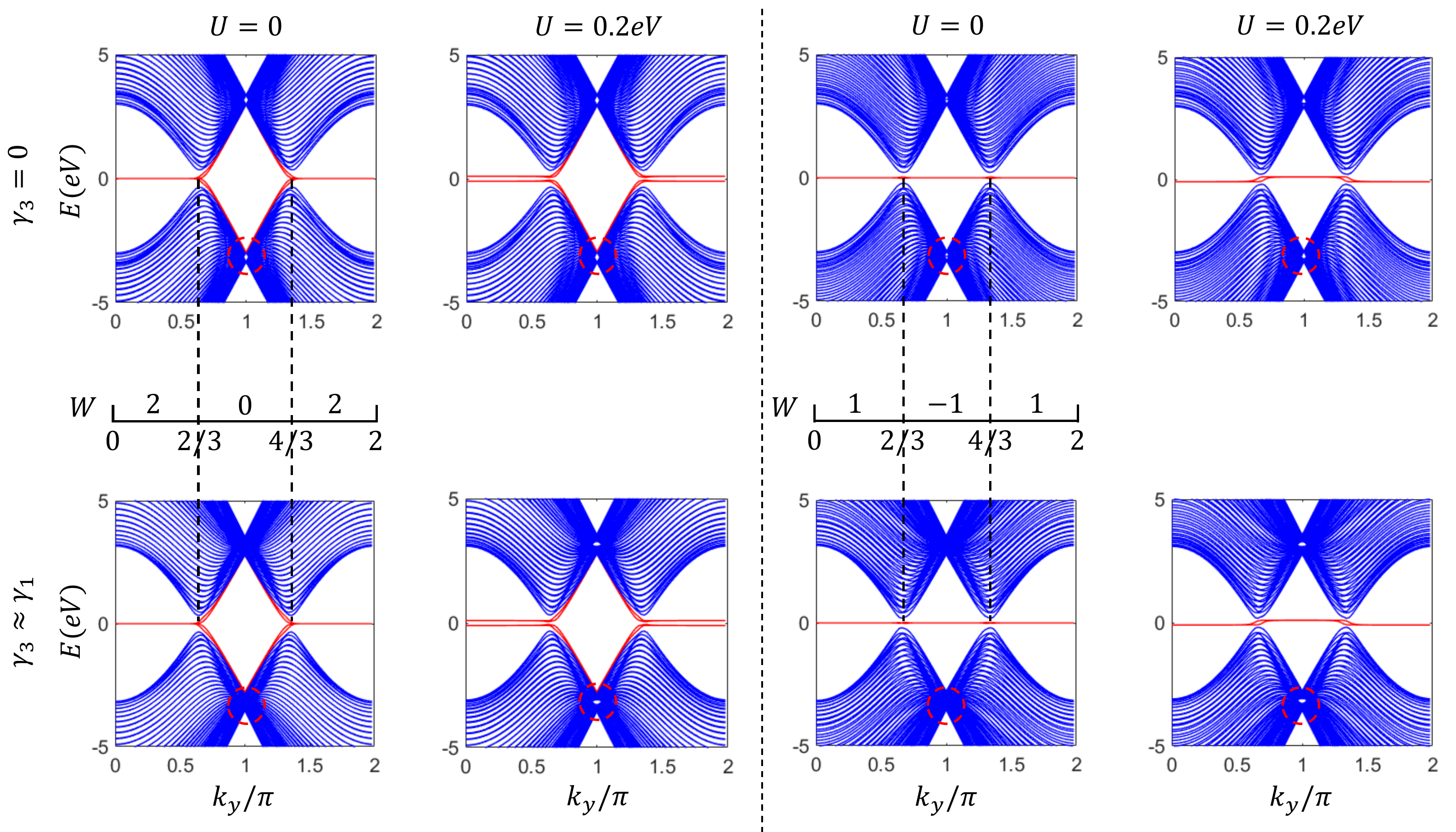}
\end{center}
\caption{(Color online) Left panel: Band structure with different interlayer
bias $U$ and NNN interaction $\protect\gamma _{3}$ corresponding to atomic
structure shown in Fig. \protect\ref{fig6}(a). Right panel: Band structure
with different interlayer bias $U$ and NNN interaction $\protect\gamma _{3}$
corresponding to atomic structure shown in Fig. \protect\ref{fig6}(d). The
red solid lines show the edge states. Related winding number $W(k_{y})$ is
shown in the middle for $U=0$ cases in both panels, where the chiral
symmetry is preserved. The red dashed circle marks the energy band area
within bulk bands where the unexpected edge states may appear, which is
discussed in detail in Fig. \protect\ref{fig9}.}
\label{fig7}
\end{figure*}

Unlike the armchair-armchair-edge case, the different geometry (distinguished by arrows) of
bearded-bearded (zigzag-zigzag) edges of BLGs do not influence the band
structure and edge states near the zero-energy (Fermi level). So we only
show the band structure of bearded-bearded nanoribbon corresponding to $H_{%
\mathrm{bea-bea}}^{\longleftarrow }\left( k_{y}\right) $ in left panel of
Fig. \ref{fig7} for simplicity. The non-trivial and trivial topological
classification straightly determines the existence of zero-energy edge
states since $U$ breaks the chiral symmetry. The zero-energy edge states
appear as flat fourfold degenerate bands when $U=0$, belonging to
non-trivial topological class. The related winding number calculation tells
that they correspond to a winding number $W=2$ when they are flat and $W=0$
when they enter the bulk bands. Flat bands from each layer are separated by
a gap when $U\neq 0$, as shown in Fig. \ref{fig7}. Bulk Hamiltonian of the
SSH ladder $H_{\mathrm{bea-bea}}(k_{y},k)(U\neq 0)$ belongs to topological
trivial class (Table. \ref{Table I}). Here the edge states for $U\neq 0$ are
still topological since they connect two topologically different Dirac
points \cite{Man}.

\vspace{-0.06cm}

\begin{figure}[tbp]
\begin{center}
\includegraphics[width=0.48\textwidth]{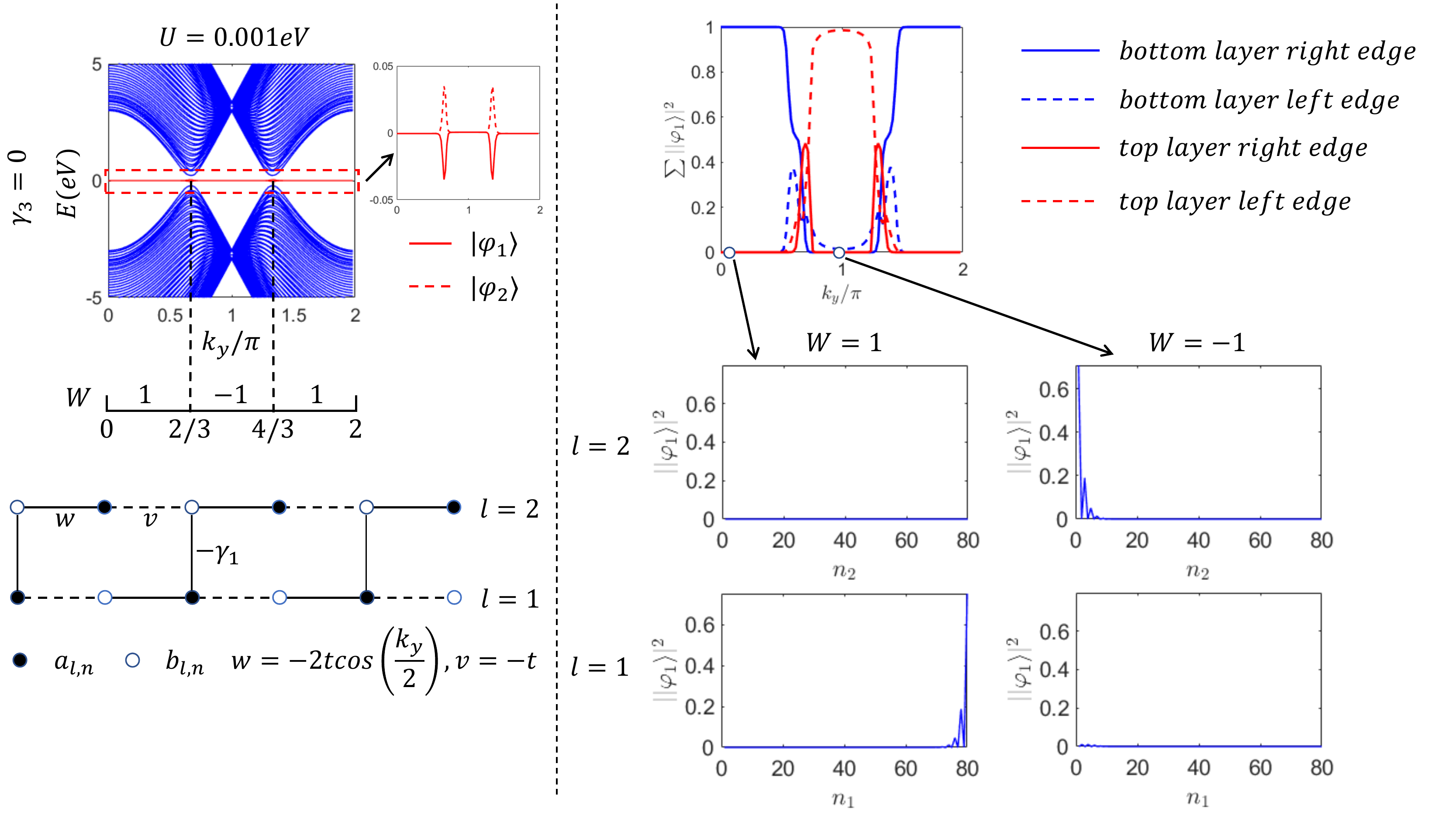}
\end{center}
\caption{(Color online) Left panel: Band structure with interlayer bias $%
U\approx 0$ and zero NNN interaction $\protect\gamma _{3}$. Corresponding
SSH ladder parameterized by $k_{y}$ is shown at the bottom. The reason for
using $U\approx 0$ is explained in the text. Right panel: The distribution
of the wave function $|\protect\varphi _{1}\rangle $ on different edges and
different layers as a function of $k_{y}$ and the distribution of the wave
function $|\protect\varphi _{1}\rangle $ in real space for two typical $%
k_{y} $ values with two different winding numbers. $n_{l=1,2}$ is site index
on the respective layer, which increases along the finite direction of the
ribbon. A transition between bottom-left and top-right is observed when
crossing the Dirac point.}
\label{fig8}
\end{figure}
\vspace{-0.06cm}

\subsection{AB-stacked BLG with bearded-zigzag edges}

Armchair MLG ribbon can only be stacked with armchair MLG ribbon to form an
AB-stacked BLG ribbon. In contrast, zigzag MLG ribbon can be stacked with
bearded MLG ribbon to form an AB-stacked BLG ribbon, which is described by
the Hamiltonian
\begin{equation}
H_{\mathrm{bea-zig(zig-bea)}}=H_{\mathrm{bea(zig)}}^{1}+H_{\mathrm{zig(bea)}%
}^{2}+H_{\mathrm{int}}+H_{\mathrm{on-site}},
\end{equation}%
Here we choose $H_{\mathrm{bea-zig}}$ as the example. The situation is the
same as what we found in above subsection, where different edge
configurations lead to similar band structures and related edge states near
the zero-energy. Hamiltonians of different edge configurations can be
denoted as $H_{\mathrm{bea-zig}}^{\swarrow \left( \nearrow \right) }\left(
k_{y}\right) $ as shown in Fig. \ref{fig6}(c) and (d), respectively. The
bulk Hamiltonian of the SSH ladder $H_{\mathrm{bea-zig}}^{\swarrow \left(
\nearrow \right) }\left( k_{y}\right) $ can be expressed as%
\begin{eqnarray}  \label{bea-zig Hamiltonian}
H_{\mathrm{bea-zig}}^{\swarrow \left( \nearrow \right) }\left(
k_{y},k\right) &=&\eta ^{\dagger }h_{\mathrm{bz}}^{\swarrow \left( \nearrow
\right) }\left( k_{y},k\right) \eta ,  \label{hbz} \\
\eta &=&\left( a_{1,k},b_{1,k},b_{2,k},a_{2,k}\right) ^{T},  \notag \\
h_{\mathrm{bz}}^{\swarrow \left( \nearrow \right) }\left( k_{y},k\right) &=&%
\left[
\begin{array}{cc}
h_{\mathrm{b}}^{1}\left( k_{y},k\right) & H_{\mathrm{int}}^{\swarrow \left(
\nearrow \right) }\left( k_{y}\right) \\
\left[ H_{\mathrm{int}}^{\swarrow \left( \nearrow \right) }\left(
k_{y}\right) \right] ^{\dagger } & h_{\mathrm{z}}^{2}\left( k_{y},k\right)%
\end{array}%
\right] ,  \notag
\end{eqnarray}%
with%
\begin{eqnarray}
h_{\mathrm{b}}^{1}\left( k_{y},k\right) &=&h_{\mathrm{b}}\left(
k_{y},k\right) -\frac{U}{2}\mathit{I}_{2}, \\
h_{\mathrm{z}}^{2}\left( k_{y},k\right) &=&h_{\mathrm{z}}\left(
k_{y},k\right) +\frac{U}{2}\mathit{I}_{2},  \notag \\
H_{\mathrm{int}}^{\swarrow }\left( k_{y}\right) &=&-\left[
\begin{array}{cc}
\gamma _{3}\left( 2e^{ik}\cos \frac{k_{y}}{2}+1\right) & 0 \\
0 & \gamma _{1}e^{ik}%
\end{array}%
\right] ,  \notag \\
H_{\mathrm{int}}^{\nearrow }\left( k_{y}\right) &=&-\left[
\begin{array}{cc}
\gamma _{1} & 0 \\
0 & \gamma _{3}\left( 2\cos \frac{k_{y}}{2}+e^{ik}\right)%
\end{array}%
\right] .  \notag
\end{eqnarray}%
When $U=0$, $H_{\mathrm{bea-zig}}^{\swarrow }\left( k_{y}\right) $ belongs
to non-trivial class $\mathcal{BDI}$. Otherwise, it belongs to trivial class
$\mathcal{AI}$ (Table. \ref{Table I}). The band structure of nanoribbon
corresponding to $H_{\mathrm{ber-zig}}^{\swarrow }\left( k_{y}\right) $ is
plotted in right panel of Fig. \ref{fig7}. The twofold degenerate
zero-energy edge state (flat band) can exist in the whole $k_{y}$ region
when $U=0$. It corresponds to a topological phase transition between two
non-trivial topological phases characterized by $W=-1$ and $W=1$ when
crossing the Dirac point, as shown in left panel in Fig. \ref{fig7}. When $%
U\neq 0$, although the chiral symmetry is broken (i.e. no well-defined
winding number), topological edge state still exists in the whole $k_{y}$
region with different energies when crossing the Dirac point. Notice that
when $U\neq 0$, the bands are no longer twofold degenerate, as explained
later.

This topological phase transition is not a numerical artifact. In contrast,
it is protected by interlayer coupling. Winding number describes the mapping
of Brillouin zone $(k)$ to $U(n)$ group, whose fundamental group is $Z$ \cite%
{Ling}. In BLG ribbon case, it is $U(2)$. This mapping is orientation
sensitive. For example, in calculating the winding number of BLG bea-bea
ribbon, as presented in Fig. \ref{fig7}. The non-zero winding number of it
can be turned into $-2$ instead of $2$ by reversing the direction of $k$. In
the BLG bea-zig case, as in the right panels of Fig. \ref{fig7}, it is the
bearded layer/zigzag layer that is responsible for the $1/-1$ part of
winding number. If we can choose the orientation of $k$ independently for each
layer, we can make the winding number always $1$, i.e. no phase transition is
present. This simple conjecture can be verified by setting $\gamma
_{3}=\gamma _{1}=0$, i.e. two decoupled ribbon. In that case, one can indeed
find a chiral operator $S$ such that the calculated winding number is
always $1$(or always $-1$). For example, it can be done by using $%
S=diag(1,-1,1,-1)$ for the Hamiltonian in Eq. (\ref{bea-zig Hamiltonian})
with $\gamma _{3}=\gamma _{1}=0$. It is permissible for an uncoupled system to
have different orientation of Brillouin zone for each subsystem. However, the presence of
interlayer coupling, which is present in the real BLG ribbon system, prohibits us from choosing the orientation of $k$
independently for each layer. We are forced to choose the same orientation
of $k$ for two layers of ribbon. Otherwise, we are not able to write $H_{int}$
in Bloch form. More specifically, this means that $S=diag(1,-1,1,-1)$ is no
longer a chiral symmetry operator for Eq. (\ref{bea-zig Hamiltonian}) with
non-zero interlayer coupling. In this sense, the observed topological phase
transition is protected by interlayer coupling.

The change in winding number can be seen from the behavior of the wave
function of edge states. For $W=1\left( \left\vert v\right\vert <\left\vert
w\right\vert \right) $ region, one of the degenerate edge states can be
approximately expressed as%
\begin{eqnarray}
\left\vert \varphi _{1}\right\rangle &\approx &\frac{1}{\sqrt{\Omega _{1}}}%
\sum_{j=1}^{N}\left( -\frac{v}{w}\right) ^{N-j}b_{1,j}^{\dagger }\left\vert
0\right\rangle ,  \label{state1} \\
\Omega _{1} &=&\frac{1-\left( v/w\right) ^{2N}}{1-\left( v/w\right) ^{2}}%
\approx \frac{w^{2}}{w^{2}-v^{2}},  \notag
\end{eqnarray}%
for $\gamma _{3}=U\approx 0$, which is equivalent to one of the edge states
of MLG ribbon with bearded edges (bottom layer), as shown in Fig. \ref{fig8}%
, where the label $l=1,2$ in $a_{l,j}^{\dagger }\left( b_{l,j}^{\dagger
}\right) $ represents the bottom/top layer respectively. The other edge
state is%
\begin{eqnarray}
\left\vert \varphi _{2}\right\rangle &\approx &\frac{1}{\sqrt{\Omega _{2}}}%
\sum_{j=1}^{N}[\left( -\frac{v}{w}\right) ^{j-1}a_{1,j}^{\dagger }\left\vert
0\right\rangle  \label{state2} \\
&&+\left( -1\right) ^{j}j\frac{\gamma _{1}}{w}\left( \frac{v}{w}\right)
^{j-1}a_{2,j}^{\dagger }\left\vert 0\right\rangle ],  \notag \\
\Omega _{2} &=&\frac{1-\left( v/w\right) ^{2N}}{1-\left( v/w\right) ^{2}}%
+\left( \frac{\gamma _{1}}{w}\right) ^{2}\sum_{j=1}^{N}j^{2}\left( \frac{v}{w%
}\right) ^{2j-2}  \notag \\
&\approx &\frac{w^{2}}{w^{2}-v^{2}}+\left( \frac{\gamma _{1}}{w}\right)
^{2}\sum_{j=1}^{N}j^{2}\left( \frac{v}{w}\right) ^{2j-2}.  \notag
\end{eqnarray}%
Since $\gamma _{1}\ll t$ (one order smaller in our choice of parameters), $%
\left\vert \varphi _{2}\right\rangle $ mainly distributes on $l=1$ layer.
When entering $W=-1\left( \left\vert v\right\vert >\left\vert w\right\vert
\right) $ region, $\left\vert \varphi _{1}\right\rangle $ and $\left\vert
\varphi _{2}\right\rangle $ continuously change to%
\begin{eqnarray}
\left\vert \varphi _{1}\right\rangle &\rightarrow &\frac{1}{\sqrt{\Omega _{1}%
}}\sum_{j=1}^{N}[\left( -\frac{w}{v}\right) ^{j-1}b_{2,j}^{\dagger
}\left\vert 0\right\rangle  \label{state3} \\
&&+\left( -1\right) ^{j}j\frac{\gamma _{1}}{v}\left( \frac{w}{v}\right)
^{j-1}b_{1,j}^{\dagger }\left\vert 0\right\rangle ],  \notag \\
\Omega _{1} &=&\frac{1-\left( w/v\right) ^{2N}}{1-\left( w/v\right) ^{2}}%
+\left( \frac{\gamma _{1}}{v}\right) ^{2}\sum_{j=1}^{N}j^{2}\left( \frac{w}{v%
}\right) ^{2j-2}  \notag \\
&\approx &\frac{v^{2}}{v^{2}-w^{2}}+\left( \frac{\gamma _{1}}{v}\right)
^{2}\sum_{j=1}^{N}j^{2}\left( \frac{w}{v}\right) ^{2j-2},  \notag
\end{eqnarray}%
and%
\begin{eqnarray}
\left\vert \varphi _{2}\right\rangle &\rightarrow &\frac{1}{\sqrt{\Omega _{2}%
}}\sum_{j=1}^{N}\left( -\frac{w}{v}\right) ^{N-j}a_{2,j}^{\dagger
}\left\vert 0\right\rangle ,  \label{state4} \\
\Omega _{2} &\approx &\frac{v^{2}}{v^{2}-w^{2}}.  \notag
\end{eqnarray}%
This indicates that when the winding number changes, the distribution of
edge states in real space will switch both edge and layer, as shown in Fig. %
\ref{fig8}.

We would like to stress that when $U$ is exactly zero, one can argue that $%
\left\vert \varphi _{1}\right\rangle $ and $\left\vert \varphi
_{2}\right\rangle $ are exponentially degenerate. The degeneracy originates
from the edge degree of freedom. Thus, any linear combination of them are
still edge states, with $\left\vert \varphi _{1}\right\rangle $ and $%
\left\vert \varphi _{2}\right\rangle $ being only one possible choice. Then
it does not make sense to discuss the behaviour of $\left\vert \varphi
_{1}\right\rangle $ in real space when crossing the Dirac point, e.g. from
bottom-left to top-right, as a linear combination would mix left edge and
right edge. The naturalness of choosing $\left\vert \varphi
_{1}\right\rangle $ and $\left\vert \varphi _{2}\right\rangle $ is seen as
following: when a small positive bias U, e.g. $0.001eV$, is included as
perturbation $H_{\mathrm{per}}$, the cross term $\left\langle \varphi
_{1}\right\vert H_{\mathrm{per}}\left\vert \varphi _{2}\right\rangle =0$.
According to the first order degenerate perturbation theory \cite{Kato}, $%
\left\vert \varphi _{1}\right\rangle $ and $\left\vert \varphi
_{2}\right\rangle $ are the good states to use in perturbation. Energy of
state in Eq. (\ref{state2}) would increase relatively to the energy of state
in Eq. (\ref{state1}), as the former has minor component on $l=2$ layer.
Similarly, energy of state in Eq. (\ref{state4}) would increase relatively
to the energy of state in Eq. (\ref{state3}), as the latter has minor
component on $l=1$ layer. Thus,$\left\vert \varphi _{1}\right\rangle $/$%
\left\vert \varphi _{2}\right\rangle $ would continuously evolve into Eq. (%
\ref{state3})/Eq. (\ref{state4}). Fig. \ref{fig8} is plotted with $U=0.001eV$
and $\gamma _{3}=0$. The behaviour of states is numerically smooth,
indicating the degeneracy is effectively lifted with a small bias.

\begin{figure*}[tbp]
\begin{center}
\includegraphics[width=0.7\textwidth]{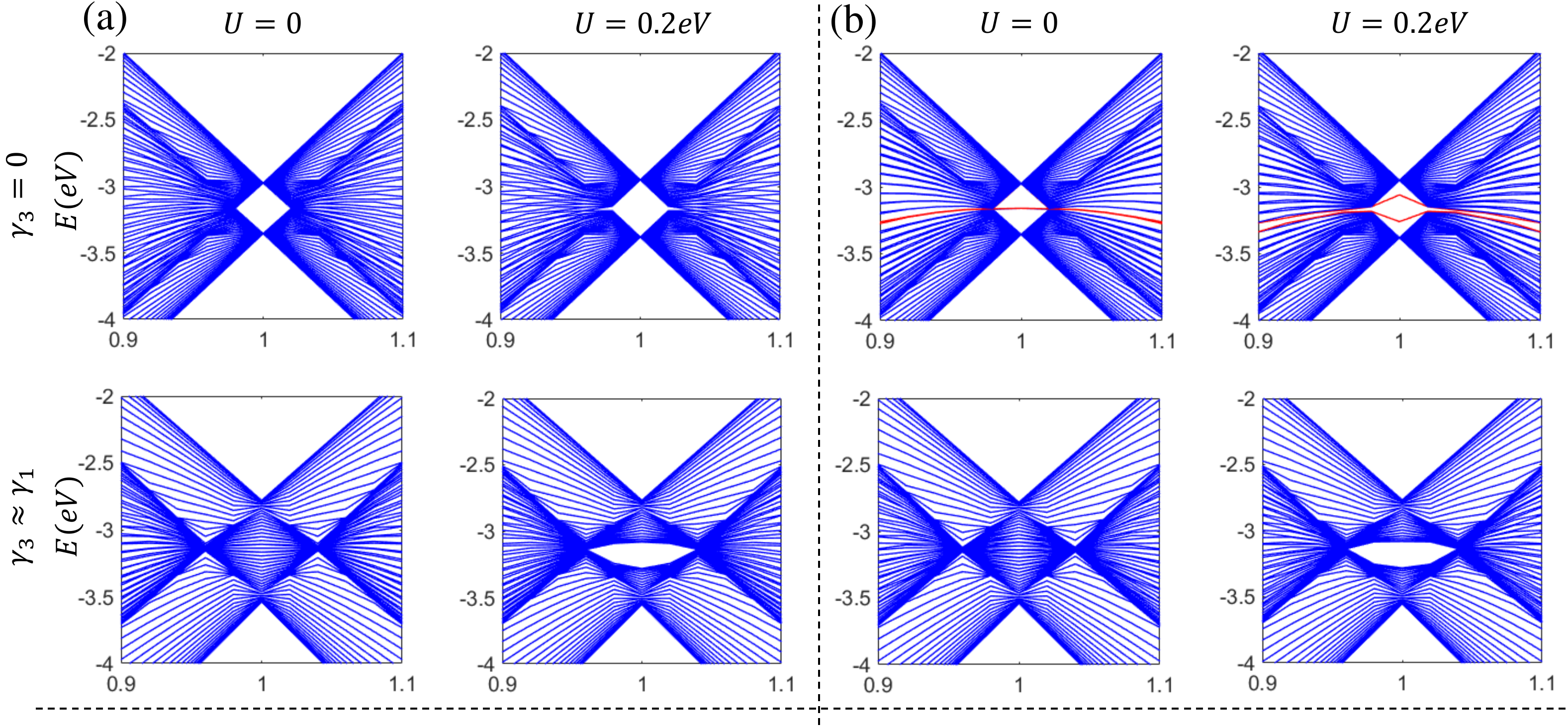} \includegraphics[width=0.7%
\textwidth]{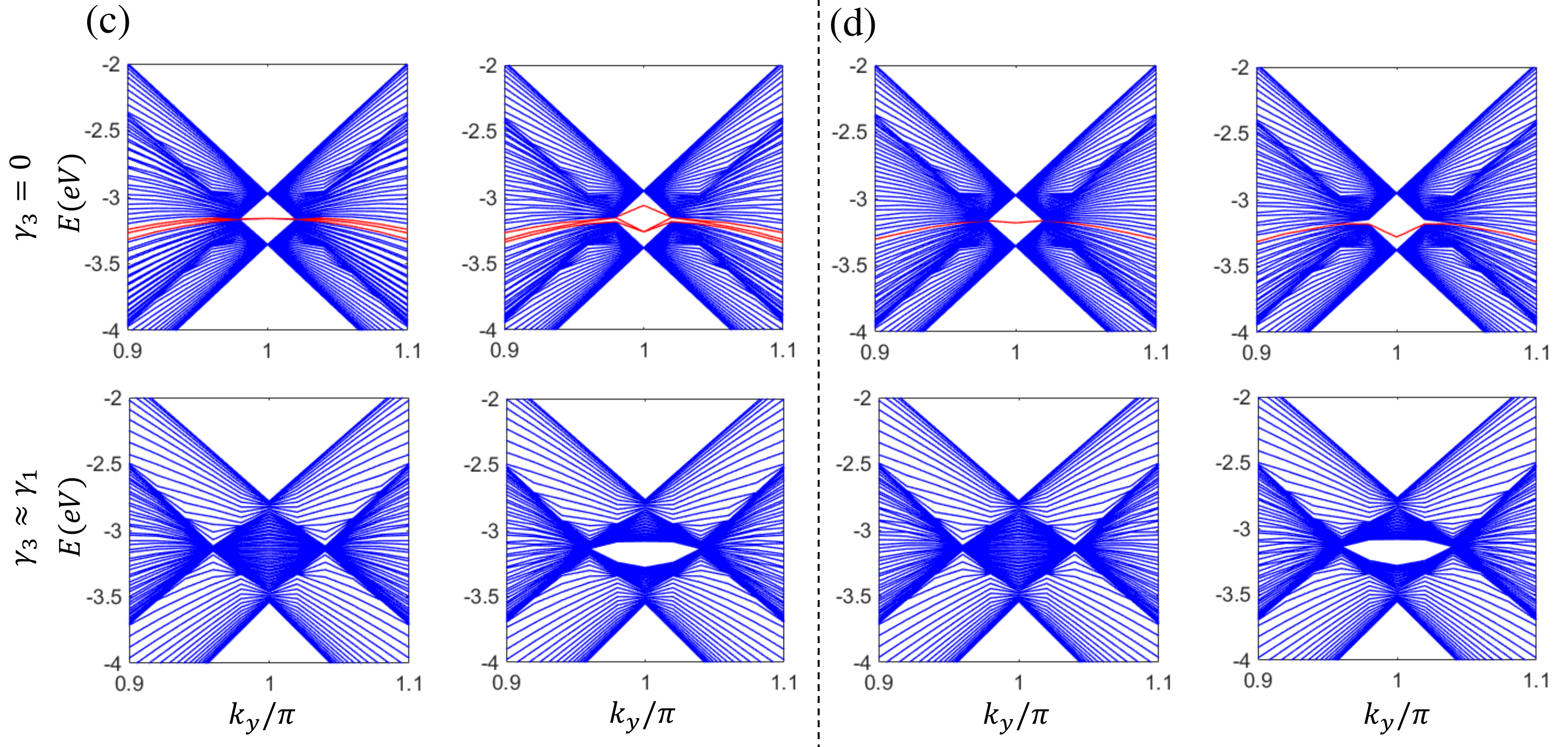}
\end{center}
\caption{(Color online) The band structures in the region marked in Fig.
\protect\ref{fig7}, coming from different edge configurations of AB-stacked
BLG ribbon with different interlayer bias $U$ and NNN interaction $\protect%
\gamma _{3}$. Only the bands with $E<0$ are shown for simplicity since $E<0$
and $E>0$ bands are symmetric. (a)-(d) correspond to the atomic structures
shown in Fig. \protect\ref{fig6}(a)-(d), respectively. The red solid lines
show the unexpected edge states appearing in the gap within the bulk bands
that are away from zero energy.}
\label{fig9}
\end{figure*}
\vspace{-0.06cm}

\begin{figure*}[tbp]
\begin{center}
\includegraphics[width=0.48\textwidth]{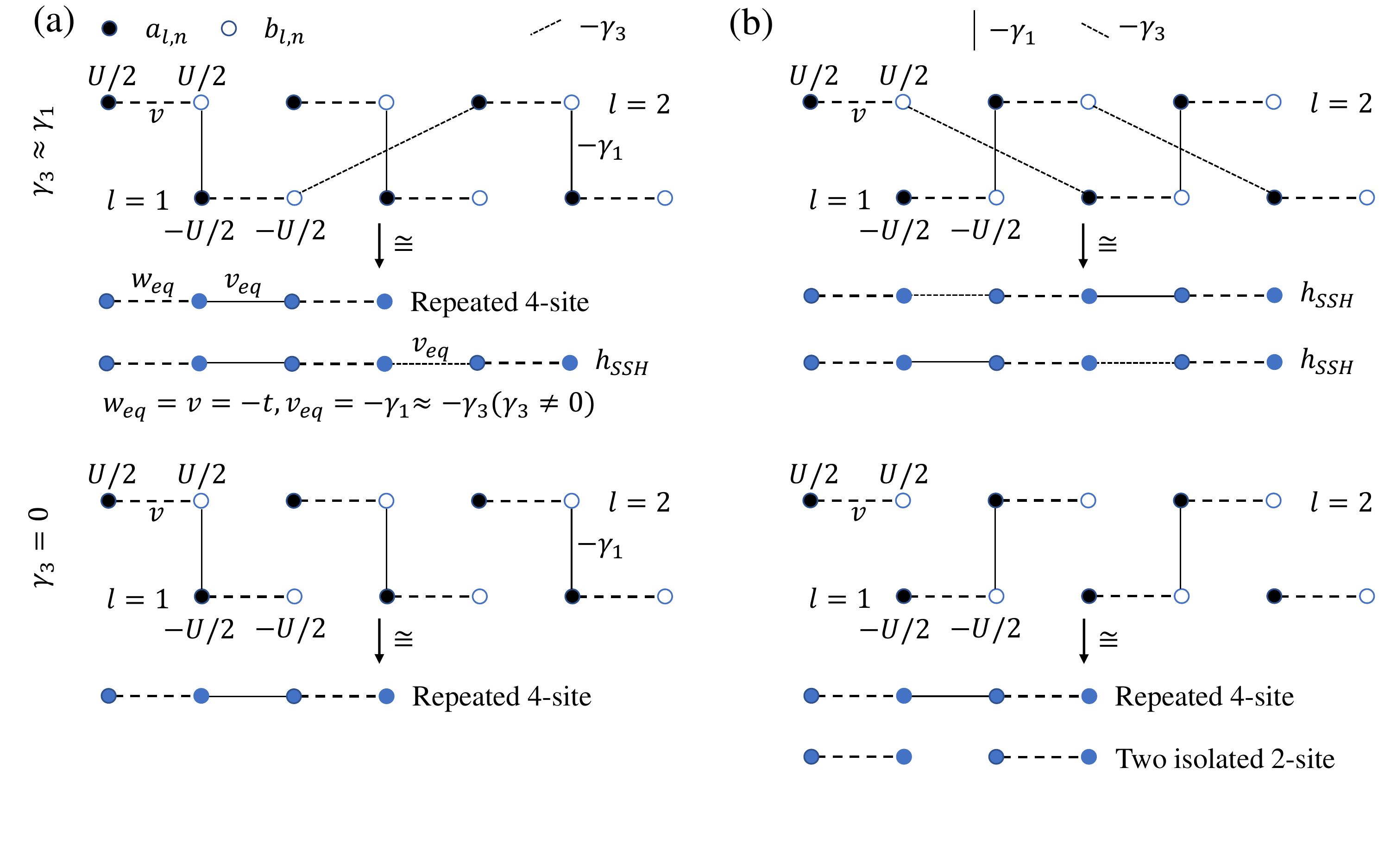} %
\includegraphics[width=0.48\textwidth]{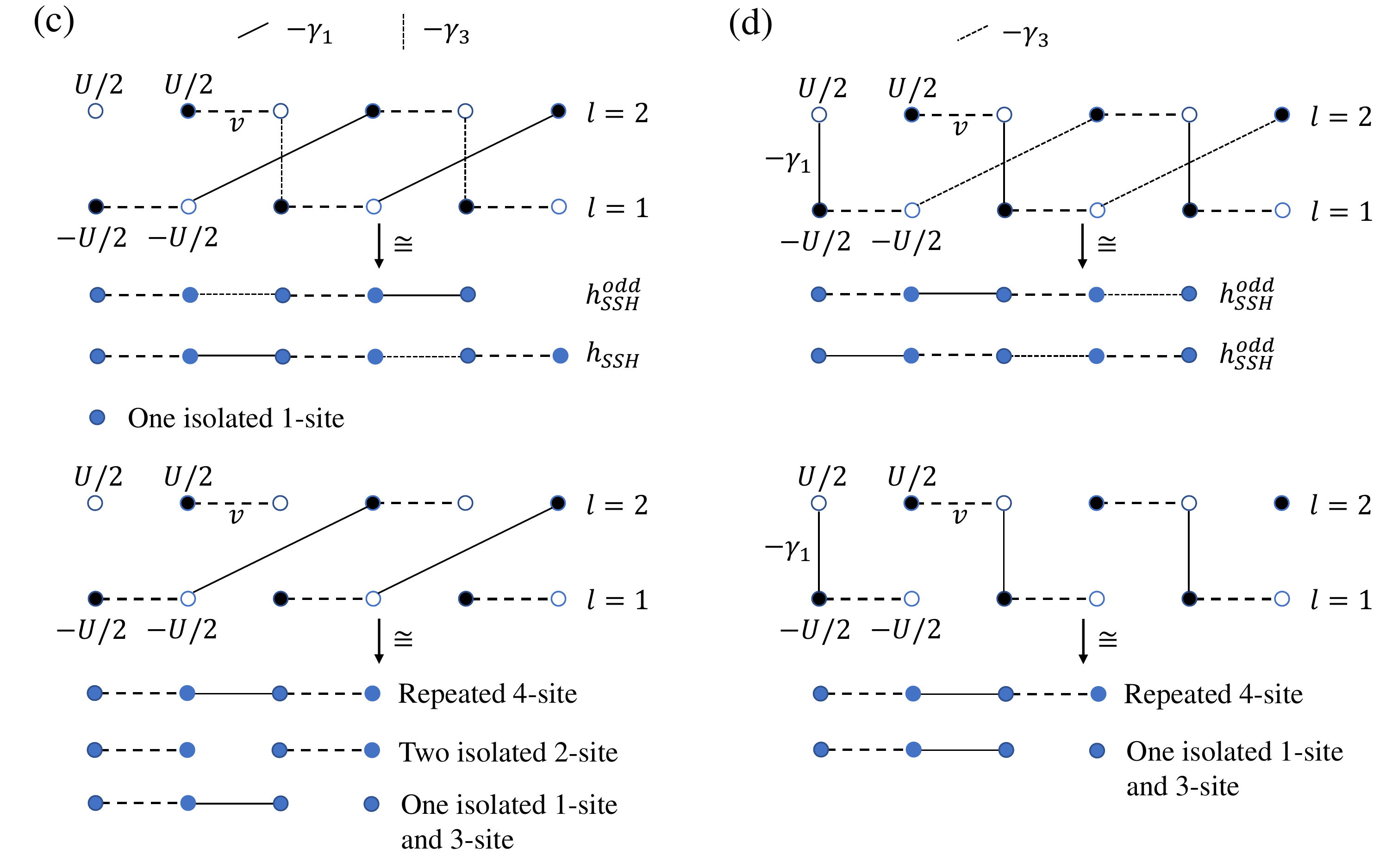}
\end{center}
\caption{(Color online) Effective SSH ladder parameterized by $k_{y}$ when $%
k_{y}\approx \protect\pi $ for different AB-stacked BLG ribbon shown in Fig.
\protect\ref{fig6}. (a)-(d) correspond to $H(k_{y})$ plotted in (a)-(d) in
Fig. \protect\ref{fig6}, respectively. The equivalent lattice structures are
shown in blue in each figure, the hopping between sites are given in (a).}
\label{fig10}
\end{figure*}
\vspace{-0.06cm}

\section{Edge states appearing in the gap away from the zero energy (Fermi
level)}

\label{Edge states appeared in the gap away from the zero energy (Fermi
level)} In above sections, we have considered two edge configurations of BLG
bea-zig ribbon and two edge configurations of BLG bea-bea ribbon, denoted
via different arrows. It was shown that, for example, two edge
configurations of BLG bea-bea ribbon have the same topology. It turns out
the difference between these two edge configurations appears in the form of
non-topological edge states, as discussed below.

Besides the topological edge states we discussed in the last section, which
exist as gapless or gapped flat bands near the zero energy (Fermi level),
some unexpected edge states are found in the gap within bulk bands that are
away from the zero energy (Fermi level) in the AB-stacked BLGs with a
bearded-bearded (zigzag-zigzag) edge or a bearded-zigzag edge, as shown in
Fig. \ref{fig9}. The existence and number of these edge states are
determined by the specific edge configurations of AB-stacked BLGs. These
unexpected edge states appear in the region near $k_{y}=\pi $ in all
situations, which reminds us their appearance may be relevant to the
effective Hamiltonians $H(k_{y}=\pi )$ for different types of edge
configurations.

Therefore, we show the lattice structure of the effective Hamiltonians $%
H(k_{y}=\pi )$ for different types of edge configurations in Fig. \ref{fig10}%
, where some couplings vanish since $\cos \left( \pi /2\right) =0$. $H_{%
\mathrm{bea-bea}}^{\longleftarrow }(k_{y}=\pi )$ with $\gamma _{3}=0$,
describes the simply repeated decoupled $4$-site structures%
\begin{eqnarray}
h_{4} &=&w_{eq}c_{1}^{\dagger }c_{2}+v_{eq}c_{2}^{\dagger
}c_{3}+w_{eq}c_{3}^{\dagger }c_{4} \\
&&+\mathrm{H.c.}+\sum_{j=1}^{2}\frac{\left( -1\right) ^{j+1}U}{2}\left(
c_{2j-1}^{\dagger }c_{2j-1}+c_{2j}^{\dagger }c_{2j}\right) ,  \notag
\end{eqnarray}%
as shown in Fig. \ref{fig10}(a). Each 4-site structure would provide 4
energy levels%
\begin{eqnarray}
E &=&\pm \frac{1}{2}\sqrt{2\gamma _{1}^{2}+U^{2}+4t^{2}\pm 2\varepsilon }, \\
\varepsilon &=&\sqrt{4U^{2}t^{2}+4t^{2}\gamma _{1}^{2}+\gamma _{1}^{4}},
\notag
\end{eqnarray}%
and each of these four energy levels are highly degenerate since there are
many identical 4-site structures. When we consider the region close to $%
k_{y}=\pi $, each of the four highly degenerate levels would split into many
bulk states according to first order degenerate perturbation theory. So
there are no edge states under this scenario in the neighbour of $k_y=\pi$
in the bulk gap.

If $\gamma _{3}\approx \gamma _{1}$ as we discussed before, the energy bands
of $H_{\mathrm{bea-bea}}^{\longleftarrow }(k_{y}=\pi )$ is dominated by not
only a simply repeated decoupled $4$-site structure but also a newly formed
SSH chain%
\begin{eqnarray}
h_{\mathrm{SSH}} &=&\sum_{n=1}^{N}w_{eq}a_{n}^{\dagger
}b_{n}+v_{eq}b_{n}^{\dagger }a_{n+1}+\mathrm{H.c.} \\
&&+\sum_{n=1}^{N}\frac{\left( -1\right) ^{n+1}U}{2}\left( a_{n}^{\dagger
}a_{n}+b_{n}^{\dagger }b_{n}\right) ,  \notag
\end{eqnarray}%
as shown in the Fig. \ref{fig10}(a). Because $t\gg \gamma _{1}$ in our
choice of parameters \cite{Kuz}, which is general in most cases \cite%
{Neto,Mc1}, this equivalent SSH chain leads to no edge states. This is in
accord with the results shown in both Fig. \ref{fig7} and Fig. \ref{fig9}(a).

However, the situation is different for $H_{\mathrm{bea-bea}}^{\swarrow
}\left( k_{y}\right) $. When $\gamma _{3}=0$, two isolated $2$-site
structures%
\begin{equation}
h_{2}=w_{eq}c_{1}^{\dagger }c_{2}+\mathrm{H.c.}\pm \frac{U}{2}\left(
c_{1}^{\dagger }c_{1}+c_{2}^{\dagger }c_{2}\right) ,
\end{equation}%
appear in addition to repeated decoupled $4$-site structures, as shown in
Fig. \ref{fig10}(b). These isolated $2$-site structures provide the
eigenstates with energy $\pm U/2\pm t$, as shown in Fig. \ref{fig9}(b). When
we consider the region close to $k_{y}=\pi $, since the energy of 2-site
structure is different from (well separated) that of bulk 4-site structure,
their eigenstates would predominantly mix among themselves instead of mixing
with states from those 4-site structures. Since these two 2-site only exist
at the edge, the mixing result would remain edge states. If $\gamma
_{3}\approx \gamma _{1}$ , two equivalent SSH chains exist as shown in Fig. %
\ref{fig10}(b). Each of these chains is the same as the structure of $h_{%
\mathrm{SSH}}$, leading to no edge states both near the zero energy (Fig. %
\ref{fig7}) and in the gap within bulk bands (Fig. \ref{fig9}(b)).

The condition is more complicated when we discuss the AB-stacked BLGs with a
bearded-zigzag edge. For $H_{\mathrm{ber-zig}}^{\swarrow }(k_{y}=\pi )$ with
$\gamma _{3}=0$, the same two isolated $2$-site structures at the end of
chain and repeated decoupled $4$-site structure appear as in previous cases.
Besides, there are also an isolated $1 $-site $h_{1}=\left( U/2\right)
c_{1}^{\dagger }c_{1}$ and a $3$-site structure at two ends of chain%
\begin{eqnarray}
h_{3} &=&w_{eq}c_{1}^{\dagger }c_{2}+v_{eq}c_{1}^{\dagger }c_{3}+\mathrm{H.c.%
} \\
&&-\frac{U}{2}\left( c_{1}^{\dagger }c_{1}+c_{2}^{\dagger
}c_{2}-c_{3}^{\dagger }c_{3}\right) ,  \notag
\end{eqnarray}%
as shown in Fig. \ref{fig10}(c). The two edge states near zero energy are
from $h_{1}$ and one of eigenstates of $h_{3}$ with close-to-zero energy.
The three edge states appearing in the gap within bulk bands shown in Fig. %
\ref{fig9}(c) are naturally described by one eigenstate of $h_{3}$ with $E<0$
and two eigenstates of two $h_{2} (U=0)$ with energy $-t$. Again, in the
neighbourhood of $k_y=\pi$, states from 1-site/2-site/3-site will mix among
themselves instead of mixing with bulk 4-site states due to energy
difference. Thus these states would remain edge states in this
neighbourhood. When $\gamma _{3}\approx \gamma _{1}$, two SSH chain
structures appear with odd number of sites, each of them provides one edge
state near zero energy \cite{Han}. All other states of odd-site SSH chain
are bulk states. Thus, there are no edge states in the gap within bulk bands
, as shown in Fig. \ref{fig9}(c).

At last we discuss $H_{\mathrm{ber-zig}}^{\nearrow }(k_{y}=\pi )$. $\gamma
_{3}=0$ leads to the repeated decoupled $4$-site structures, one isolated $1$%
-site, and one $3$-site structures as shown in Fig. \ref{fig10}(d). The edge
state shown in Fig. \ref{fig9}(d) comes from one eigenstate of $h_{3}$ with $%
E<0$. The two near-zero edge states are eigenstate of $h_{1}$ and one of
eigenstates of $h_{3}$. If $\gamma _{3}\approx \gamma _{1}$, two SSH chains
structures appear with odd number of sites. Together they provide two
zero-energy edge states, and no edge states in the gap within bulk bands, as
shown in Fig. \ref{fig9}(d).

As a conclusion, number of edge states away from Fermi-energy is determined
by the number of isolated structures when $k_y=\pi$, which is in turn
dependent upon the edge configuration of AB-stacked BLGs.

\section{Conclusions and discussions}

\label{Conclusions and discussions}

In this paper, we discussed the existence and topology of edge states in
AB-stacked BLG ribbon with various edge configurations. We illustrated the
correspondence between BLG ribbon and SSH ladder. A detailed topological
classification based on discrete symmetry and topological invariants
calculation for effective 1D bulk Hamiltonian of SSH ladder $H\left(
k_{y},k\right) $ parameterized by $k_{y}$ of AB-stacked BLG ribbon were
constructed,showing the bulk-edge correspondence between zero-energy edge
states of the ribbon and winding number of $H\left( k_{y},k\right) $.

In addition, we found the existence of bias-induced edge states in the
armchair-armchair BLG ribbon. This is not topologically protected, as
discussed in the text. However, this is still an interesting result, as it
implies that edge states can be produced via a mechanism that does not have
anything to do with the edge of systems\cite{footnote}. We also found a topological phase
transition between two topologically non-trivial phases in zigzag-bearded
BLG ribbon, corresponding to a twofold degenerate zero-energy edge state
existing in the whole $k_{y}$ region. We demonstrated that when a edge state
crosses the phase transition point, it will switch both layer and edge.

Moreover, we pointed out that some non-topological edge states without the
protection of the chiral symmetry can be found in the gap within bulk bands
that are away from zero-energy (Fermi level). The existence and number of
these states are sensitive to the edge configurations of BLG ribbons even if
their bulk topologies are the same, which can be simply explained by
effective Hamiltonians $H(k_{y}=\pi )$ for different situations. Though we
focus only on the honeycomb lattice in this paper, it should be obvious that
our study can be generalized to lattices of different shapes, such as Kagome
\cite{Tian,Ni,Nir} and triangular \cite{Kuru} lattices, and of higher
dimensions, such as the description of edge states and surface states in
three dimensional topological insulators \cite{Teo,Shan}. All of these
provide potential directions for further study.

\section*{Acknowledgments}

T. X. Tan would like thank Z. A. Hu and F. R. Fan for useful discussions. C.
Li would like to thank B. Fu for insightful suggestions. The work is support
by the University Grants Committee/Research Grant Council of the Hong Kong
SAR (AoE/P-701/20), the HKU Seed Funding for Strategic Interdisciplinary
Research, and the Croucher Senior Research Fellowship.

\appendix*
\section{The difference between wave vector number $k_y$ and $k$}

The following example is provided to make the discussion in section \ref{Edge
states in the MLG with different edges} of
the main text more concrete. Consider the ribbon depicted in Fig. \ref{fig1}(a), the
Hamiltonian is given by Eq. (2) of the text:

\begin{eqnarray}
H_{bea} &=&-t\sum_{m=1}^{M}\{\sum_{n=1}^{N}\left[ a_{m,n}^{\dagger
}b_{m,n}+b_{m,n}^{\dagger }(a_{m,n+1}+a_{m+1,n+1})\right]   \notag \\
&&-b_{m,N}^{\dagger }(a_{m,1}+a_{m+1,1})+H.c.\}
\end{eqnarray}

For this specific ribbon, $N=5$. The value of $M$ is unimportant since the
ribbon is infinite (periodic) along that direction. It is a common practice
to assume that there is a periodicity in the operator, i.e. $%
a_{i,j}^{(\dagger )}=a_{i+M,j+N}^{(\dagger )}$ and $b_{i,j}^{(\dagger
)}=b_{i+M,j+N}^{(\dagger )}$. Notice that under this condition, the
Hamiltonian given by Eq. (2) is invariant when adding $1$ to all the first
subscript of all operators. But it is not invariant when adding $1$ to all
operators's second subscript. This difference is due to the term $%
b_{m,N}^{\dagger }(a_{m,1}+a_{m+1,1})$, hence we say that this term gives
the open boundary condition.

By basic solid state physics, we are allowed to make Fourier transformation
along the invariant direction, which is the direction of the first
subscript. The Fourier transformation is given by Eq. (3) of the main text,
while the resulting Bloch form Hamiltonian is given by Eq. (4). Consider now
doing the following redefinition of operators, which is always allowed since
it does no affect the band structure.
\begin{eqnarray}
a_{k_{y},n}^{\dagger } &\rightarrow &a_{k_{y},n}^{\dagger }e^{ik_{y}n/2}, \\
b_{k_{y},n}^{\dagger } &\rightarrow &b_{k_{y},n}^{\dagger }e^{ik_{y}n/2}
\notag
\end{eqnarray}

After this redefinition, the Hamiltonian in Eq. (4) will be transformed into
the form of Eq. (5), which is the following, where $a_{n}$/$b_{n}$ are
shorthand for $a_{k_{y},n}$/$b_{k_{y},n}$
\begin{equation}
H_{bea}(k_{y})=\sum_{n=1}^{4}(va_{n}^{\dagger }b_{n}+wb_{n}^{\dagger
}a_{n+1})+va_{N}^{\dagger }b_{N}+H.c.
\end{equation}

\begin{figure}[tbp]
\begin{center}
\includegraphics[width=0.48\textwidth]{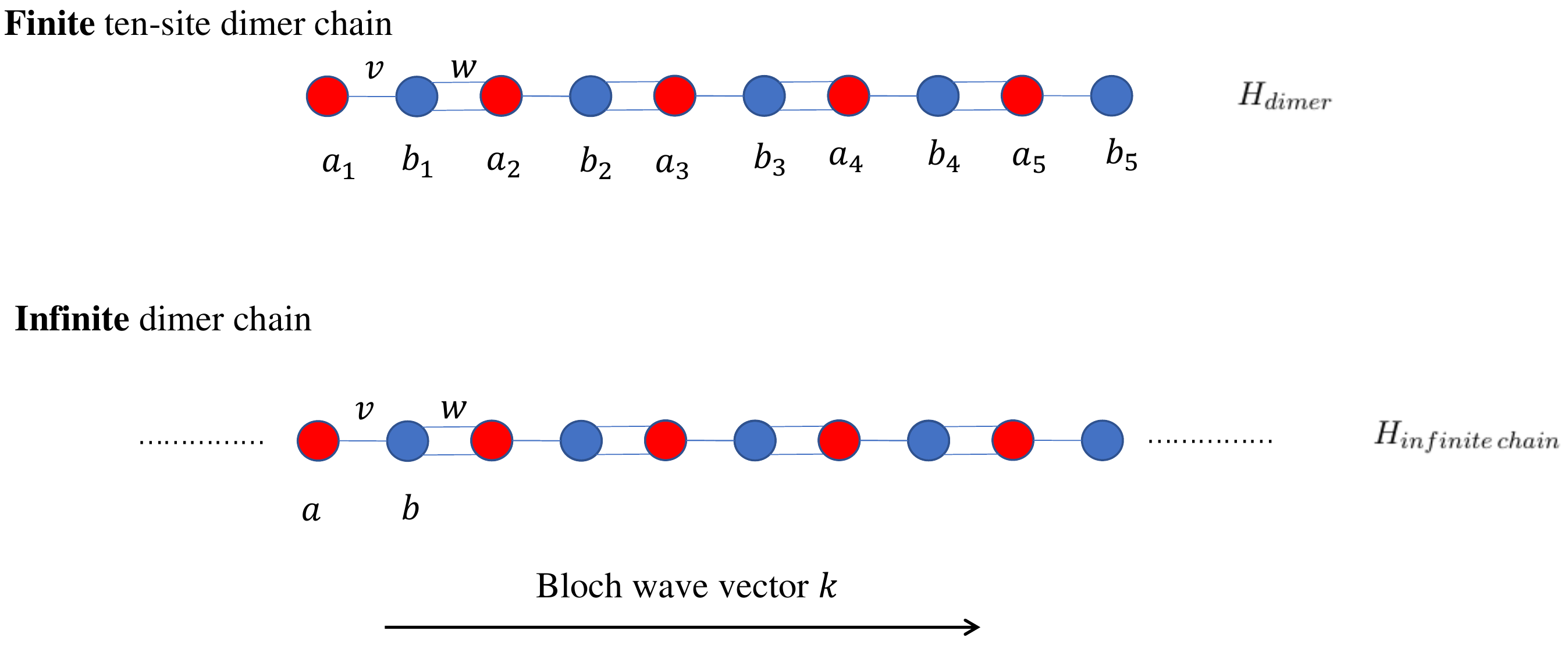}
\end{center}
\caption{(color online)}
\label{figa1}
\end{figure}

From this point on, we will leave MLG ribbon, and instead consider a $10$%
-site SSH chain as illustrated in Fig. \ref{figa1}. Its intercell coupling $w
$ and intracell coupling $v$ are defined as following:
\begin{equation}
v\equiv -t\quad w\equiv -2tcos(\frac{k_{y}}{2})
\end{equation}

Readers should refrain from associating $k_{y}$ appearing above with the one
obtained in the Fourier transformation of MLG ribbon, but instead should
think it just as a parameter on which $v$ and $w$ depend. The Hamiltonian of
the $10$-site SSH chain is:
\begin{equation}
H_{\mathrm{dimer}}=\sum_{n=1}^{4}(va_{n}^{\dagger }b_{n}+wb_{n}^{\dagger
}a_{n+1})+va_{N}^{\dagger }b_{N}+H.c.
\end{equation}%
$H_{\mathrm{dimer}}$ is formally equivalent to $H_{\mathrm{bea}}(k_{y})$,
i.e. $H_{\mathrm{dimer}}\cong H_{\mathrm{bea}}(k_{y})$. However, it is more
obvious that this is an additional trick we can play with this $10$-site SSH
chain. We can make this chain infinite and obtain a Hamiltonian $H_{\mathrm{%
chain}}$ that can be put in Bloch form, with corresponding Bloch wave vector
$k$, as used in the main text.
\begin{equation}
\begin{split}
& H_{\mathrm{infinite\,chain}}=\sum_{k\in FBZ}H(k) \\
& H(k)=\left( v+we^{-ik}\right) a_{k}^{\dagger }b_{k}+H.c. \\
& =%
\begin{pmatrix}
0 & v+we^{-ik} \\
v+we^{ik} & 0%
\end{pmatrix}%
\end{split}%
\end{equation}%
This $H(k)$ is dependent on $k_{y}$ through $v$ and $w$. This $H(k)$ is what
is referred to as $h_{b}(k_{y},k)$ in Eq. (6) of the text, with which we can
calculate winding number $W$ for different $k_{y}$ using Eq. (7) and Eq. (8)
of the text. For this specific example, the chiral operator $S$ is
\begin{equation}
S=%
\begin{pmatrix}
1 & 0 \\
0 & -1%
\end{pmatrix}%
\end{equation}%
This procedure can be easily generalized to bilayer case as we use in later
sections.

\begin{thebibliography}{99}
\bibitem{footnote} The presence of these edge states is still dependent on edge configuration, since only one of the two edge configurations of BLG arm-arm ribbon show them. However, the mechnism itself, i.e. bias, does not treat the atoms on the edge differently from those deep inside the bulk. This is different from edge states induced by modulting only the edge onsite energy. More specifically, after we impose this bias, the effective $k_y-parameterized$ SSH ladder for BLG arm-arm ribbon still has a bulk Hamiltonian, on which we can discuss the topology. However, if we only modulate the onsite energy on the edge of BLG arm-arm ribbon, the effective SSH ladder does not have a bulk Hamiltonian.
\end{thebibliography}
\end{document}